\documentclass[preprint,aps]{revtex4}
\usepackage{graphicx}
\usepackage{bm}
\usepackage{color}
\usepackage{amssymb}
\usepackage{epsfig}

\begin{document}
\topskip 20mm

\title{Random matrix ensembles with column/row constraints: part I}
\author{Pragya Shukla and Suchetana Sadhukhan}
\affiliation{Department of Physics, Indian Institute of Technology,
Kharagpur, India}

\date{\today}

%\widetext

\begin{abstract}
% .

	We  analyze statistical properties of a complex system subjected to conditions which manifests  through specific constraints on the column/row sum of  the matrix elements of its Hermitian operators. The presence of additional constraints besides real-symmetric nature leads to new correlations among their eigenfunctions, hinders a complete delocalization of dynamics and affects the eigenvalues too.  The statistical analysis of the latter indicates the presence of a new universality class analogous to that of a special type of Brownian ensemble appearing between Poisson and Gaussian orthogonal ensemble.

\end{abstract}

\pacs{  PACS numbers: 05.40.-a, 05.30.Rt, 05.10.-a, 89.20.-a}

\maketitle

\section{Introduction}

Linear operators with fixed sum-rules on the columns/rows of their matrix elements  appear in  widely different areas e.g. disordered systems {\color{red}\cite{kirk, gc, bp, bcm, mm, gmpv, ps,scs,rmo,pin,ach,sm,zs,gw, gmpuv, os, yvf, luck, bray0, adrs}}, complex networks \cite{shep}, financial markets \cite{cm} etc. Missing information due to  complexity leads to randomization of the operator and  it can appropriately be represented by a random matrix which satisfies all system specific conditions. The statistical behavior of the operator can then be modeled by a multi-parametric random matrix ensemble, with each of its matrices subjected to a fixed column/row sum-rule. A  sub-class of such matrices, known as stochastic matrices or Markov matrices have been studied in past; almost all these studies focused on the  properties of individual eigenvalues and eigenfunctions \cite{mar}. In  context of a complex system however  such an information does not serve much purpose: eigenvalues, eigenfunctions as well as other physical properties fluctuate from sample to sample and even within one sample and 
a knowledge of their average behavior is not sufficient. This motivates us to pursue an statistical analysis of  the eigenfunctions and eigenvalues of the random matrix ensembles with column/row sum rule.

 The conditions influencing the nature of a matrix ensemble can broadly be divided into two types. The "global" or "matrix" constraints e.g., symmetry or conservation laws  which affect the nature of each matrix i.e its transformation and structural properties and introduce collective relations among the elements. On the contrary, the "local" or more appropriately "ensemble"  constraints manifest themselves through  ensemble parameters i.e the distribution properties of the matrix elements. A  "matrix" constraint  e.g. column sum-rule can coexist with  different combinations of the "ensemble" constraints; this gives rise to the possibility of different random matrix ensembles with same matrix  constraint.  It is therefore desirable to understand not only the influence of matrix constraints on the statistical fluctuations but also the role played by the ensemble constraints. The present study considers the effect of a combination of global constraints e.g. Hermiticity and time-reversal symmetry besides column/row sum rule, as well as the ensemble constraints (e.g. disorder), on the matrix ensembles. 
Systems with such constraints exist in diverse areas e.g bosonic Hamiltonians such as  phonons, and spin-waves in Heisenberg and XY ferromagnets, antiferromagnets, and spin-glasses,  euclidean random matrices, random reactance networks, financial systems and Internet related Google matrix  etc.

	The paper is organized as follows. Before proceeding for the mathematical analysis of column constrained matrices, it is natural to query the origin of these constraints from a physicist's perspective.   
The section II briefly introduces 
a few examples from different areas so as to reveal wide-applicability of the ensembles with column/row constraints.  The section also helps to reveal connections among seemingly different areas in which a same mathematical constraint originates from different physical conditions. The column/row constraints introduce new correlations among the matrix elements which influence  their distribution and can lead to a wide range of random matrix ensembles; section III discusses  how various possibilities may arise in context of the constrained matrices with real-symmetric elements. The eigenvalues/eigenfunction fluctuations being standard tools to analyze the ensemble statistics, it is natural to seek the imprints of constraints on them. The section IV analyzes
the effect of these constraints on the joint eigenvalue-eigenfunction density. The latter turns out to be analogous to that of a well-known ensemble, known as Brownian ensemble, without column/row constraints and with many known, although approximate,  results for its fluctuation measures;   the analogy  is used in section V to gain statistical information about the column constrained ensembles (CCE).   The section VI concludes this study with a brief reviews of our main results and open questions.

.

\section{Examples of Systems with column/ row sum rules}

The appearance of column/row constraints in the matrices representing a complex system is not directly obvious. 
To explain, we briefly discuss  few examples from different areas.

\vspace{0.1in}

\noindent{\bf  Goldstone modes:} 
Goldstone modes are  low-energy excitations in a system (e.g. phonons, spin-waves)  in which a continuous symmetry (also referred as Goldstone symmetry) of the Hamiltonian is broken by the ground state. 
The Goldstone symmetry represents the invariance of a linear operator, say Hamiltonian $H$,  under a uniform shift in a local configuration variable. This implies $\left[H, b \right]=0$ where $b= \sum_{k=1}^N b_k$, with $b_k$ as the  creation operator for the shift at the basis-state $|k \rangle$ in a discrete $N$-dimensional basis ($k=1,..,N$). The mode $\sum_{k=1}^N |u_k \rangle$, with $|u_k\rangle = b_k |0\rangle$ is then an eigenstate of $H$, with the same energy $E_0$ as the ground state $|0 \rangle$ however it may differ in response to an external perturbation (thus indicating lack of symmetry).

The appearance of column/row constraints in systems with Goldstone modes can be explained as follows \cite{kirk}. 
Using the $2nd$ quantization form $H=\sum_{m,n} V_{mn} \; b_m^{T} b_n$ and the relation
$i \frac{\partial b_k}{\partial t}=\left[H, b_k \right]$, equation of motion for the state $|u_k(t)\rangle$ with $|u_k(t)\rangle = {\rm e}^{-i \omega t} |u_k\rangle$  becomes 
\begin{eqnarray}
\omega \; u_l =\sum_{k=1}^N V_{kl} \; u_k
\label{vkl}
\end{eqnarray}
with $\omega=E-E_0$ as the excitation energy and $E_0$ as the ground state energy.  The uniform shift  $u_1=u_2=..=u_N$ in the ground state ($\omega=0$)  then gives the "column constraint" 
$\sum_{k=1}^N \; V_{kl} =0$.

Eq.(\ref{vkl}) is  the eigenvalue equation for the matrix $H = V$ with eigenvalue $\omega$ and eigenfunction $u$ with $u_k \equiv \langle k|u\rangle$, $k=1,\ldots,N$ as its components. The excitation spectrum can then be obtained by an exact diagonalization of $V$ subjected to zero column constraint. As bosonic excitations are characterized by $\omega > 0$, this subjects $V$ to additional constraints (i) $V$ must also be semi-positive definite, (ii) the eigenvector corresponding to lowest eigenvalue should be delocalized in the basis where the column/row constraint is satisfied. Eq.(\ref{vkl}) can also be rewritten as the eigenvalue equation for another matrix $\mathcal{H}$, defined by $E \; u_l =\sum_{k=1}^N \mathcal{H}_{kl}  \; u_k$, with $\mathcal{H}_{kl} = V_{kl}+ E_0 \; \delta_{kl}$. $\mathcal{H}$ is therefore subjected to non-zero column constraint  $\sum_{k=1}^N \; \mathcal{H}_{kl} =E_0$. Clearly the eigenvalues  of $\mathcal{H}$ are same as those of $V$ except for a constant shift $E_0$. The spectral statistics of the two matrices is therefore analogous. But the analogy does not extend  to their eigenfunction statistics.

\vspace{0.1in}

\noindent{\bf  Random euclidean matrices:} A $N \times N$ euclidean random matrix, say $H$, represents an operator associated with a random distribution of $N$ points, with its entries given by a deterministic function of the distances between the points \cite{mm}. Consider $N$ points, characterized by position $x_i$  in a $d$ dimensional euclidean space, then  
\begin{eqnarray}
H_{ij} = f(x_i-x_j) - u \; \delta_{ij} \sum_k f(x_j -x_k)
\end{eqnarray}
with $u$ as a real parameter.  The case $u=1$  satisfies the column/row constraint $\sum_{k=1}^N H_{kl}=0$; this case appears e.g in  studies on vibrational properties of amorphous solids (glasses or supercooled liquids) \cite{mm, gmpv} or random master equation \cite{bray0}. The case $u=0$ corresponds to the cases with euclidean correlations among matrix elements which subjects matrix elements to additional constraints besides column/row constraints.  (The stability requirement of matter imposes another constraint i.e spectrum to be positive definite). Such matrices exist in many  areas e.g random lasing or nonlinear optical phenomena \cite{ps}, collective spontaneous emission (super-radiance) \cite{scs}, multiple scattering or waves in random medium and Anderson localization  \cite{rmo,pin,ach} and instability in nonlinear random medium \cite{sm, zs, gw}.

\vspace{0.1in}

\noindent {\bf Conductance in  nano-wire junctions and random reactance networks} 
Natural appearance of the multiple wire-junctions in any quantum circuit has motivated great deal of research interest  and different frameworks have been introduced to probe their transport properties. In Tomonaga-Luttinger (TL) model of a quantum wire-junction, the  scattering matrix $S$ at the junction can be expressed in terms of a  Hermitian matrix $U$ which is subjected to column/ row constraints  due to Kirchoff's laws along with other conservation laws \cite{bcm, adrs}. For example, for a junction of $N$ wires, each point $Q$ in the bulk can be parameterized by 
the pair $(x,i)$ with $E_i$ as the edge, $i=1\ldots N$, and $x$ as the distance of point $Q$ from the vertex along that edge. As discussed in \cite{bcm},  with a point like defect localized at the vertex of the junction,  both $S$ and $U$ are $N\times N$ matrices: 
\begin{eqnarray}
S(k)=- \frac{\lambda (1-U) - k(1+U)}{\lambda (1-U) + k (1+U)}
\label{sk}
\end{eqnarray}
with $k$ as the conjugate variable to $x$  and $\lambda$ as a parameter which fixes the scale at which scattering matrix is given exactly by $U$: $S(k=\lambda)=U$ for $\lambda \not=0$. 
As discussed in \cite{bcm}, both $S$ and $U$ are subjected to the constraint $\sum_{m=1}^N S_{ml} = \sum_{m=1}^N U_{ml}=1$; here $m, l=1 \dots N$ label the edges of the junction. As  
The column/ row constrained matrices also appear in  random RLC networks, made of random mixture of capacitance $C$, inductance $L$ and resistance $R$ \cite{yvf, luck}; here again the constraint arises  due to Kirchoff's laws. 

\vspace{0.1in}

\noindent {\bf Page Rank Algorithm and Google Matrix:} 
The information retrieval from the enormous database of world Wide Web (WWW) is based on various algorithms to rank the web-page. One such algorithm, known as page rank algorithm (PRA) is the basis of Google search engine; it efficiently determines a vector, referred as page rank vector,  ranking the nodes of a network by order of their importance. 
This vector is an eigenfunction of the Google matrix \cite{shep} which is related to the adjacency matrix $A$ of the complex network connecting the nodes of world-wide-web. 

Assuming $N$ nodes, $A$ can be written as a $N \times N$ matrix, with its elements characterizing the links between the nodes. Another matrix $S$, corresponding to the transitions in a Markov chain of the network, can now be  constructed from $A$ such that the sum of all elements in any column of $S$ is equal to unity:
$S_{ij}={A_{ij} \over \sum_k  A_{kj}} $ if $\sum_k  A_{kj} > 0$ and $S_{ij}=\frac{1}{ N}$ if $A_{kj}=0$ for all $k=1,2,.. N$. Such a construction replaces the columns, with zero matrix elements (referred as dangling nodes) by a constant value $1/N$; this adds a link from every dangling node to every other node and is suitable for PageRank algorithm. The Google matrix can now be written as 
\begin{eqnarray}
G = \gamma \; S + (1-\gamma) E 
\end{eqnarray}   
where  $E$ is a uniform matrix: $E_{ij}=1$ and $\gamma$ is known as a damping factor (with $(1-\gamma)$ corresponding to a surfer's probability to jump to any page).
As clear from the definition, both $S$ and $G$ satisfy the column sum-rule $\sum_i S_{ij} =1$ and 
$ \sum_i G_{ij} =\gamma + (1-\gamma) N$.

\vspace{0.1in}

\noindent {\bf Financial markets and pattern games}: 
Pattern games are well-studied realistic agent-based models of financial markets. The dynamic nature of interactions among  economic agents   
(due to constant thinking and altering the decision) gives rise to fluctuations which are similar in nature as in a disordered system  (although their  origins are different).  Several attempts have been made in past to describe pattern games as  disordered spin systems \cite{cm}; the 
available information about the disordered-spin dynamics 
can  further be used to probe the market-dynamics.  Similar ideas  can be extended  to conceive a game of  interacting agents with multiple strategies which can be mapped to a classical spin system with Goldstone modes; these ideas are yet to be explored.

\section{Effect of column/row sum rule on matrix elements distribution }

Consider a complex system with its behavior described by a $N \times N$ real-symmetric random matrix $H$ subjected to column/row constraints:
\begin{eqnarray}
\sum_{k=1}^N H_{kl} =\sum_{k=1}^N H_{lk} = \alpha_l.
\label{gs}  
\end{eqnarray}
where $\alpha_l$ is a constant;
the rows being same as columns in this case, eq.(\ref{gs})  will be referred hereafter as "column constraints" only and $\alpha_l$  as "column constant".

 %(note, $H$ being real-symmetric, the constraints on rows are same as those on columns and therefore will not be mentioned in further discussion).
%eq.(\ref{gs}) can also be expressed as a row constraint $\sum_{k=1}^N H_{lk} =\alpha_l$).
%
In absence of any other information available about the system, the probability density  $\rho(H)$ of the matrix elements of $H$ can be derived using information-theoretic concepts:  among all possible distributions subjected to given constraints, $\rho(H)$ is given by the one which minimizes the amount of information or alternatively maximizes the information entropy 
(known as maximum entropy hypothesis).
A standard measure of the amount of information carried by a distribution $\rho(H)$ is given by Shannon's entropy, defined as 
\begin{eqnarray}
I[\rho(H)] = - \int \rho(H)\; {\rm ln}\; \rho(H)\; {\rm d}\mu(H),
\label{ent}
\end{eqnarray}
with ${\rm d}\mu(H)$ as the measure in $H$-space. Using $\delta I =0$ along with the column constraints and any other known constraints, the mathematical form of $\rho(H)$ can  be derived using Lagrange multipliers method (see page 61 of \cite{me} for detailed discussion).

As an example, we consider the case
  (i) $\rho(H)$ is normalized;  
(ii) the mean $<H_{kl}> \equiv \int  \; H_{kl} \; \rho(H) \;  {\rm d}\mu(H)$ and correlations $<H_{ij} H_{kl}> \equiv \int H_{ij} H_{kl} \; \rho(H) \; {\rm d}\mu(H)$ of the off-diagonals are known, 
(iii) the column constraint is given by eq.(\ref{gs}). These conditions  
alongwith the condition $\delta I =0$  then leads to Gaussian form of $\rho$ (\cite{me}):
\begin{eqnarray}
\rho(H) = \rho_s(H)  \; \;
\prod_{l=1}^N \delta \left(\sum_k  H_{kl} - \alpha_l \right) .
\label{rho}
\end{eqnarray}
where $\rho_s(H)$ describes the probability density of the off-diagonals 
\begin{eqnarray}
\rho_s(H) = \mathcal{N} \; \prod_{l=1}^{N}  {\rm exp}\left[-  (\mathcal{H}_l-\mathcal{M}_l )^{T} \;  ( C^{(l)} )^{-1} \;  (\mathcal{H}_l-\mathcal{M}_l)\right]  .
\label{rhos}
\end{eqnarray}
with $\mathcal{N} $ as a normalization constant, $\mathcal{H}_l= \left[H_{kl}\right]$ and $\mathcal{M}_l  =\left[ \; \langle H_{kl} \rangle \;  \right]$ as the column vectors of size  $N-1$, consisting only of the off-diagonals and their mean, respectively. 
Further $C^{(l)}$ is the $N \times N$ covariance matrix,  with its elements
$C_{kj}^{(l)} \equiv \langle H_{kl} \; H_{jl} \rangle -\langle H_{kl} \rangle \langle H_{jl} \rangle $  describing the correlations between the elements of vector $H_l$ (note the variance $\sigma^2_{kl} =C_{kk}^{(l)})$. The ensemble density (\ref{rho}) with $\rho_s$ given by eq.(\ref{rhos}) is later referred as the column constrained  Gaussian ensemble (CCGE).

A  Gaussian form of $\rho_s$ in eq.(\ref{rhos}) results only 
from the known constraints on  the $1^{\rm st}$ and  $2^{nd}$ order moments of each off-diagonal.  The column constraint is imposed in eq.(\ref{rho}); it determines the diagonal density 
from those of the off-diagonals:
\begin{eqnarray}
\langle H_{ll} \rangle  &=& \alpha_l - \sum_{k=1; k\not=l}^N \langle H_{kl} \rangle 
 \label{cons1} \\ 
\langle H_{ll}^2 \rangle &=&  \alpha_l^2  + \sum_{k, j=1; \not=l}^N \langle H_{kl} \; H_{jl} \rangle - 2 \alpha_l \sum_{k=1; k\not=l}^N \langle H_{kl} \rangle 
 \label{cons2}
\end{eqnarray}
with $\langle . \rangle$ implying ensemble averaging. 
 Information about higher order, individual moments can in general result in a non-Gaussian ensemble density. 
For example,  the maximum entropy distribution turns out to be  bimodal  if each off-diagonal  is independent and can take two possible values e.g. $H_{kl} =\pm a$, with $a$ as a constant: 
\begin{eqnarray}
\rho_s(H) =  \prod_{k,l;  k < l}^{N} \left[ \delta(H_{kl} -{a}) + \delta(H_{kl} + {a} ) \right]
\label{birho}.
\end{eqnarray}
  Eq.(\ref{rho}) with $\rho_s$ given by eq.(\ref{birho}) is later referred as the column constrained  bimodal ensemble (CCBE).   Note such distributions are often considered in spin-glass studies and random euclidean matrices; see for example \cite{stinch1}.

Here we mention only the  constraints leading to the Gaussian and bimodal densities, the two  being often used distributions for the  numerical as well as mathematical analysis.  The system may also be subjected to constraints other than those on the matrix elements moments e.g  eigenvalue density etc.   The distributions for a few other cases are discussed in \cite{me, pbera}.

\section{Effect of column sum-rule on eigenvalues/ eigenfunctions}

The sum rules on the entries of $H$ in a column  give rise to  constraints on its eigenvalues and eigenfunctions which in turn influence their distributions. We first analyze the effect on  a single matrix and subsequently on the ensemble. Here, for clear presentation of our ideas, we  confine our study to  the case with same column constant for all columns: $\alpha_l = \alpha$ for all $l=1 \rightarrow N$.

%$\sum_{k=1}^N H_{kl} =\sum_{k=1}^N H_{lk}  = \alpha$,

\subsection{Behavior for a single  matrix}
 
Consider a $N \times N$ real-symmetric, column-constrained  matrix $H$ with $E$ and $O$ as its eigenvalue and eigenfunction matrices, respectively;  their elements can be given as $E_{mn} =e_n \delta_{mn}$ and $O_{kn}$ where $O_{kn}$ refers to the $k^{th}$ component of the eigenfunction $O_n$ corresponding to eigenvalue $e_n$ (note $O_{kn}$ is real in the basis $|k \rangle$ in which $H$ is real-symmetric). 
%
%The real-symmetric nature of $H$ requires $O $,  to be orthogonal matrix: $O^{T}. O=I$ with $I$ as the identity matrix. Thus, 
%
As for a real-symmetric matrix without column constraints,  the eigenfunctions in this case are mutually independent too, with their  components distributed around a unit circle (i.e $\sum_{k} O_{kn} O_{km} = \delta_{mn}$ due to $O$ being orthogonal).

%Using the eigenvalue equation $H O = O E$, eq.(\ref{gs}) manifests itself in form of constraints on the eigenfunctions and eigenvalues of $H$;  

The eigenvalues $e_n$ are given by the  roots of the characteristic polynomial ${\rm Det}(\left( H- E  \right)$.  Invoking  the sum-rule  $\sum_{k=1}^N H_{kl} = \alpha$ and  noting that the addition of a scalar multiple of one row to another  does not change the  determinant,  it is easy to show that one of the rows, say the first one, of the  determinant ${\rm Det}\left( H- \alpha I  \right)$ is zero. As a consequence, 
\begin{eqnarray}
{\rm Det } \;\left( H-\alpha I  \right) = 0.
\label{dh}  
\end{eqnarray}
which in turn  implies   $\alpha$ as one of the eigenvalues of $H$. 

The eigenvalue equation $H O = O E$ gives the relation $\sum_{k,l=1}^N  H_{kl} O_{ln} =  \sum_{k=1}^N O_{kn} e_n$. Using column sum rule, this  can be rewritten as    
\begin{eqnarray}
 (e_n-\alpha) \; \sum_{k=1}^{N}  O_{kn} = 0  \qquad       n=1,...,N
\label{egc}  
\end{eqnarray}
which implies
\begin{eqnarray}
\sum_{k=1}^{N}  O_{kn} = 0   \qquad \qquad  e_n \not= \alpha
\label{oz}
\end{eqnarray}
Squaring eq.(\ref{oz}), rearranging terms and using the normalization condition further gives 
\begin{eqnarray}
\sum_{k,l=1; k\not= l}^N O_{kn} O_{ln} = -\sum_{k=1}^N O^2_{kn} =-1  \qquad  {\rm for} \quad e_n \not= \alpha.
\label{egc+}  
\end{eqnarray}

The condition for the eigenvector components corresponding to eigenvalue $\alpha$ can be derived as follows. Using eq.(\ref{gs}), 
one can write 
 ${\rm Tr} \; H =N \alpha - \sum_{k,l; k\not=l} H_{kl}$. This  alongwith the relation $H_{kl} = \sum_{n=1}^N e_n O_{kn} O_{ln}$ and  ${\rm Tr} \; H=\sum_{n=1}^{N} e_n = \sum_{n,k =1}^{N} e_n \; O_{kn}^2 $ gives 
\begin{eqnarray}
\sum_{n=1}^{N} e_n \; \sum_{k,l=1}^{N}  O_{ln} \; O_{kn}  =N \alpha
\label{egc1}  
\end{eqnarray}
Substitution of  eq.(\ref{egc}) in the above equation leaves,  on its left side, only  the term corresponding to non-random eigenvalue $\alpha$, thus  reducing it as
\begin{eqnarray}
\left(\sum_{k}^{N}  O_{kn} \right)^2=N    \qquad  {\rm if} \quad e_n = \alpha.
\label{egc1}  
\end{eqnarray}
The above gives $\sum_{k=1}^{N} O_{k n} = \pm \sqrt{N}$  for $e_n= \alpha$ which alongwith   normalization condition $\sum_{k=1}^N O_{kn}^2=1$ implies $O_{kn} =1/\sqrt{N}$ with $k=1 \rightarrow N$ (or  $O_{kn} =-1/\sqrt{N}$ for all $k$). The eigenvector corresponding to the non-random eigenvalue  is therefore uniformly extended in the whole basis-space. (As all other eigenvectors must be orthogonal to the uniform eigenvector, this reconfirms the sum rule (\ref{oz}) ).

As clear from eq.(\ref{oz}),  an eigenvector corresponding to $e_n \not=\alpha$  can not be localized just to one basis-state i.e $O_{kn} \not=\delta_{kn}$; it must be spread over at least a pair of them (referred henceforth as the "pair-wise localization").  This indicates a new characteristic of a typical eigenvector (except the uniform eigenvector) of a column constrained matrix: the lack of single basis-state localization which is in constrast with an unconstrained real-symmetric matrix. As expected, this affects the maximum possible value of the inverse participation ratio (IPR), the standard tool to describe the localization behavior of an eigenvector and defined as $I_2(O_n)= \sum_{k=1}^N  |O_{kn}|^4$. It is easy to see that $ I_2 \le \frac{\kappa}{2}$ with $\kappa=1$ or $2$ for a typical  eigenvector of a matrix with or without column constraints, respectively.

It is worth noting here that the column constant $\alpha$ determines the non-random eigenvalue  but   does not enter the eigenfunction constraints eq.(\ref{oz}) and eq.(\ref{egc1}).  As the non-random eigenvalue can be made zero by a shift of the origin of the spectrum, a real-symmetric matrix with column constant $\alpha \not=0$ is equivalent to $\alpha=0$. But the column constants are expected to play a more important role in the case in which they vary from one column to the other (eq.(\ref{gs}) with different $\alpha_l$); this case is still under study.

\subsection{Distribution of the eigenvalues-eigenfunctions}

Our next step is to derive the joint densities of the eigenvalues and/or eigenfunctions from the ensemble density given by eq.(\ref{rho}). As the objective here is to understand the effect of column constraints on the statistics, $H$ is assumed to have  no symmetry constraints which could result in degenerate eigenvalues (except accidental degeneracy). $H$ therefore has only one eigenvalue  equal to $\alpha$.  (Note, for the degenerate case, the matrix can be written in a block form in the symmetry-preserving basis; the present analysis is then applicable to each block). As this can  be  any one of the $N$ eigenvalues,  hereafter we choose  $e_N=\alpha$ (without loss of generality). For later reference, the eigenvector constrains can be rewritten as  
\begin{eqnarray}
\sum_{k=1}^{N} O_{k n} = \pm \sqrt{N} \; c_n,  \qquad \qquad \sum_{k,l=1; k\not=l}^{N}  O_{l n} \; O_{k n} = (N c_n-1),
\label{egc3}  
\end{eqnarray}
with $c_N=1, c_n=0$ for $n \not= N$, along with $O_{kN}= \frac{1}{\sqrt{N}}$ for all $k=1 \rightarrow N$. 

Eq.(\ref{rho}) alongwith the relation $P(E;O)= \rho(H) J(E,O|H)$, with $J(E,O|H)$ as the Jacobian of transformation from the $H$-space to 
$(E, O)$ space gives the joint density $P(E,O) \equiv P({e_1,\ldots, e_N;  O_{11}, O_{12}, \ldots ,O_{NN}})$ of the $N$ eigenvalues $e_n$ and corresponding eigenfunction components $O_{mn}$ with $m, n= 1,\ldots N$   (as defined above, $E, O$ refer to the matrices of eigenvalues and eigenvectors). Note the matrix elements constraints now transform to those on the eigenvalues-eigenfunctions):
\begin{eqnarray}
P(E;O) &=& \rho_s(E; O) \; \; J(E,O|H) \;\; \delta(e_N-\alpha) \; \; \delta_o(O)
\label{t1}
\end{eqnarray}
where $\rho_s(E; O)$ is obtained from $\rho_s(H)$ by substituting $H=O^{\dagger} E O$: $\rho_s(E;O) \equiv \rho_s(O^T E O)$. 
and the $\delta_o(O)$ represents the constraints on the eigenfunctions:
\begin{eqnarray}
\delta_o(O) \; \equiv  \;  \prod_{n=1}^{N} \delta\left( \sum_{k=1}^N  O_{kn} -\sqrt{N} c_n\right) \; \delta(O^{\dagger} O -I )
\label{delo}
\end{eqnarray}

To proceed further, a knowledge of $J(E,O|H)$ is necessary. The column constraints along with Hermitian condition gives the $J(E,O|H)$ as  

\begin{eqnarray}
J(E, O|H) =\Delta_{N}(E) \mid_{e_N=\alpha} \; = \;  \prod_{k,l=1; k < l }^{N-1} \; \mid e_k-e_l \mid  \; \prod_{n=1}^{N-1} \; \mid e_n-\alpha \mid 
\label{J} 
\end{eqnarray}
with  $\Delta_{N}(E) \equiv \Delta_N(e_1, e_2, \ldots, e_N)$ as the Jacobian of a real-symmetric matrix without column constraint:
\begin{eqnarray}
\Delta_{N}(E)  \equiv   \prod_{k,l=1; k \not=l}^{N} \; \mid e_k-e_l \mid.
\label{Delta}
\end{eqnarray}

An integration of eq.(\ref{t1}) over $O$-space, with ${\rm D}O$ as the Haar-measure,  
gives the joint eigenvalue density $ P_e(e_1,...e_N)$: 
 \begin{eqnarray}
P_e(e_1, e_2,.., e_N) 
&=&  \delta(e_N-\alpha)  \; \Delta_{ N}(E)   \;   \int \;    \;  \rho_s(E; O) \; \delta_o (O) \; {\rm D}O 
\label{pe1}
\end{eqnarray} 
The integration in eq.(\ref{pe1}) depends on the form of $\rho_s$ and is in 
general technically complicated. As examples, here we consider some simple 
cases:

\vspace{0.2in}

\noindent{\bf Case I:  $N=2$}: The integration in eq.(\ref{t1}) is  straightforward if $H$ is a $2\times 2$ real-symmetric  matrix with column constant $\alpha$. This can be shown as follows. The column constraint along with  real-symmetric nature of $H$ implies $H_{11}=H_{22}=\alpha-H_{12}=\alpha-H_{21}$.  The density  
$\rho(H)$ can then be written as 
\begin{eqnarray}
 \rho(H) &=& \rho_s(H_{11}) \delta(H_{11}+H_{12}-\alpha) \delta(H_{22}+H_{12}-\alpha)
\label{rho0}
\end{eqnarray}

The eigenvalues of $H$ are 
$e_1=H_{11}+H_{22}-\alpha =2 H_{11}-\alpha$ and $e_2=\alpha$ (thus only $e_1$ is random). This along with eq.(\ref{rho0}) and $ J=\mid e_1-e_2 \mid$ gives  
\begin{eqnarray}
  P_e(e_1, e_2) 
 = {C}\; \mid e_1-\alpha \mid \; \rho_s((e_1+e_2)/2) \; \delta(e_2-\alpha)
\end{eqnarray}
with $C$ as a normalization constant. It is easy to derive the distribution $P(s)$ of the eigenvalue-spacing $s \equiv (e_1-e_2)$, the only relevant spectral statistics in this case:  $P(s)= |s |\; \rho_s(\alpha + s/2)$.

\vspace{0.2in}

\noindent{\bf Case II:  Independent, Gaussian distributed off-diagonals with zero mean and same variance }

\vspace{0.2in}

Next we consider the ensemble density given by eq.(\ref{rho}) with  
\begin{eqnarray}
\rho_s(H) &=& \mathcal{N} \;   {\rm exp}\left[-\gamma \sum_{k,l; k\not= l} H^2_{kl} \right] 
\label{rho2}
\end{eqnarray}
with $\gamma$ arbitrary; this case is later referred as the infinite range CCGE case or $d=\infty$ CCGE case. 
A preliminary idea about the expected statistics can be developed by first looking at the relative moments of the matrix elements.   
As here $\langle H_{kl} \rangle =0$, $\langle H_{kl} H_{jl} \rangle = {1\over 2 \gamma} \; \delta_{kj}$, eqs.(\ref{cons1}, \ref{cons2}) give the first two moments of the diagonals as (with $\alpha_l = \alpha$ for all $l$)
$\langle H_{ll} \rangle = \alpha,  \quad \langle H_{ll}^2 \rangle  = \alpha^2 + {N-1 \over 2 \gamma}$.
A typical diagonal is therefore very large as compared to a typical off-diagonal, the ratio of their variance given as  
 \begin{eqnarray}
\frac{ \langle H_{ll}^2 \rangle - \langle H_{ll} \rangle^2} {\langle H_{kl}^2 \rangle - \langle H_{kl} \rangle^2} = {N-1}.
 \label{cons4}
\end{eqnarray}
This indicates the tendency of a typical eigenfunction to localize in the basis-space. But the weaker strength of  the off-diagonals is compensated by their large number  ($=N(N-1)$) and the localization is expected to be weak.

The above idea is further strengthened by looking at the eigenvalue-eigenvector statistics.  
Using the relation $H=O^{T} E O$, one can show $\sum_{k,l; k\not=l} H^2_{kl} = \frac{1}{2} \sum_{m,n=1}^{N} |e_m-e_n|^2 \; f_{mn}$ (see appendix A for the derivation). This on substitution  in eq.(\ref{rho2}) gives 
\begin{eqnarray}
\rho_s(H) = \mathcal{N}   \;
{\rm exp}\left[- \frac{\gamma}{2} \; \sum_{m, n=1}^{N} |e_m-e_n|^2 \;  f_{mn} \right]. 
\label{pe4}
\end{eqnarray}
Here $\mathcal{N}$ is the normalization constant: $\mathcal{N}= \left(\frac{2 \gamma}{\pi}\right)^{N(N-1)/4}$
and  $f_{mn}$ is a measure of correlations between the eigenfunctions $O_m$ and $O_n$:
\begin{eqnarray}
f_{mn}(O) =f_{nm}(O) &=&\sum_{k=1}^N O_{km}^2 O_{kn}^2. 
\label{fmn}
\end{eqnarray}
Thus $f_{mn} \approx 0$ if  $O_m, O_n$ are localized on different  basis states, $f_{mn} \approx 1/N$ if both are  extended almost in  whole basis. 
Further $f_{mN}=f_{Nn}=1/N$ (due to $O^2_{kN}=1/N$ for all $k$ alongwith the normalization condition $\sum_{n=1}^N O^2_{kn}=1$). 
By separating the repulsion terms of type $|e_m-e_N|$ from those of type $|e_m-e_n|$,  eq.(\ref{pe4}) can be rewritten as 
\begin{eqnarray}
\rho_s(H) = \mathcal{N}   \;  {\rm exp}\left[
-  {\gamma} \sum_{m=1}^{N-1} (e_m-\alpha)^2 \;  f_{mN}
-  {\gamma\over 2} \sum_{m, n=1}^{N-1} |e_m-e_n|^2 \;  f_{mn} \right].  
\label{pe4+}
\end{eqnarray}
As clear from the above, the repulsion of all other eigenvalues from the fixed one acts like a confining potential.  
Substitution of eq.(\ref{pe4+})  in eq.(\ref{pe1}) for $\rho_s(H) \equiv \rho_s(E;O)$ gives the joint eigenvalue  density $P_e(e_1, e_2,.., e_N) $ for the ensemble (\ref{rho2}).
Using a transformation of variables $r_n={\sqrt{\gamma}} \; (e_n-\alpha)$, $P(e_1,e_2,..,e_N)$ can also be expressed in $\alpha, \gamma$-free form

\begin{eqnarray}
P_r(r_1, r_2,.., r_N) 
&=&    \int \; P_c (r;O) \; \delta_o(O) \;  {\rm D}O 
\label{pe5}
\end{eqnarray}
where
\begin{eqnarray}
P_c(r;O) &=& \mathcal{N}   \; \delta(r_N) \;\Delta_{N}(r) \; {\rm exp}\left[
- \sum_{m=1}^{N-1} r_m^2 \;  f_{mN}
-  \frac{1}{2}\sum_{m, n=1}^{N-1} |r_m-r_n|^2 \;  f_{mn} \right].  
\label{pe5+}
\end{eqnarray}
The above implies an independence of the  statistics from the column constant  $\alpha$ or the distribution parameter $\gamma$. Further as $f_{mN} =\frac{1}{N}$, this  reduces eq.(\ref{pe5+}) to 
\begin{eqnarray}
P_c(r;O) 
&= &  \mathcal{N}   \; \delta(r_N) \; \Delta_{N}(r)  \; 
{\rm exp}\left[-\frac{1}{N} \;  \sum_{m=1}^{N-1} r_m^2 -\frac{1}{2} \;    \sum_{m, n=1}^{N-1} \; |r_m-r_n|^2 \; f_{mn} \right]  
\label{pe5a}
\end{eqnarray}

The joint eigenvalue-eigenvector density given by eq.(\ref{pe5a}) 
is derived from eq.(\ref{rho2}) without any approximation.
To proceed further, an integration of eq.(\ref{pe5}) over $O$-space in needed but  the basis-dependence of $f_{mn}$ makes it  technically difficult; (contrary to unitary space, the results for an integration over orthogonal space are still not known; see \cite{itz, gor}).  Important  insight can however be gained by following qualitative analysis (based on a similar idea used in \cite{pich}).
As clear from eq.(\ref{pe5a}), the behavior of $P_c$  depends on the competition between two sums in the exponent. 
The first term chiefly acts as a confining potential on the mutually repelling eigenvalues;  it makes $P_c(r;O)$  very small for $ |r_n| \gg \sqrt{N}$, implying the spectrum support to be of the order $\sqrt{N}$. 
The second term dictates the degree of localization of the eigenfunction and level-repulsion of the eigenvalues; as clear, $P_c(r;O)$ is large for the cases when $N \; |r_m-r_n|^2 \;  f_{mn}  \ll 1$. 
Thus the eigenfunctions $O_m, O_n$  corresponding to  large energy separations $|r_m-r_n|$ are more probable to have small overlap and  tend to occupy different parts of basis-space ($f_{mn} \rightarrow 0$ for $|r_m-r_n| > N^{-1/2}$).  But those at short energy-ranges may share the same space (for $|r_m-r_n|  < N^{-1/2}$, the 2nd term in the exponent is negligible even if $f_{mn}$ is maximum i.e $f_{mn} \approx \frac{1}{2}$).  The combined effect of the two terms is therefore an increased level-density in the bulk (for $|e_n |\ll \sqrt{N}$) with eigenfunctions relatively more extended as compared to those in the edges.   A typical $O$-matrix for the ensemble (\ref{rho2}) then consists of localized, partially localized as well as extended eigenfunctions.  Writing    $O = S +A$ where $S$ is diagonal and $A$ antisymmetric, one gets, for $m \not=n$,  
\begin{eqnarray}
f_{mn} = (S_{nn}^2 + S_{mm}^2) \; A_{mn}^2 + 
\sum_{k=1}^N \; A^2_{km} \; A^2_{kn}
\label{fmn1}
\end{eqnarray}
For a pair $O_m, O_n$ ($m,n \not=N$) occupying different parts of the basis-space (if $A_{kn} \not= 0$ then $A_{km} \rightarrow 0$), the last term of eq.(\ref{fmn1}) is relatively small. 
This gives 
\begin{eqnarray}
f_{mn} \; \approx \; (S_{nn}^2 + S_{mm}^2) \; A_{mn}^2, \qquad {\rm for }\; \;  m,n \not= N
\label{fmn2} 
\end{eqnarray}
Note the above approximation is also applicable to those $O_m, O_n$ pairs in which one of them is localized and other extended. 
The $2^{nd}$ sum in the exponent can then  be separated in two parts, one corresponding to uncorrelated $O_m, O_n$ pairs (those for which eq.(\ref{fmn2}) is applicable) and the other containing contributions from rest of the eigenfunctions; the two parts will be referred as $\sum_{uncor}$ and $\sum_{rest}$ respectively.  
Using now orthogonal space Haar measure ${\rm D}O = \prod_{n=1}^{N} {\rm d}S_{nn} \; \prod_{k< n} {\rm d}A_{kn}$, eq.(\ref{pe5a}) can  be integrated over  $A_{mn}$ variables with subscripts $m,n$ referring to  uncorrelated eigenfunction-pairs.  Being a Gaussian integral, this eliminates the repulsion terms $|r_m -r_n|$ corresponding to uncorrelated $O_m, O_n$ pairs and gives 
\begin{eqnarray}
P_r(r_1,\ldots, r_N) \approx \mathcal{N}    \; \delta(r_N)  \; 
\prod_{m<n; m,n \in rest} |r_m-r_n | \;  \prod_{k=1}^{N-1} \; r_k \;
{\rm exp} \left[-\frac{1}{N} \;  \sum_{m=1}^{N-1} r_m^2 \right] \; I(\{r \}) 
\label{pe5b}
\end{eqnarray}
where
\begin{eqnarray}
I(\{r \}) = \;\int  
\left[\prod_{m,n \in uncor} \frac{\sqrt{\pi}}{\sqrt{S_{nn}^2+S_{mm}^2} } \right]\;      {\rm exp}\left[ - \sum_{rest} \; |r_m-r_n|^2 \; f_{mn} \right]
 \; \prod_{n=1}^{N} {\rm d}S_{nn} 
\; \prod_{k< n; k \not=m} {\rm d} A_{kn} \nonumber \\
\label{Ir}
\end{eqnarray}
Further insight in eq.(\ref{pe5b}) can be gained by noting  
that a typical term in the $\sum_{rest}$ is $O\left(\frac{1}{N^2} \right)$ ($|r_m-r_n|$ now very small and $f_{mn} \sim \frac{1}{N}$ for rest of the eigenvector-pairs) and its contribution is insignificant as compared to other exponent in eq.(\ref{pe5b}). Approximating ${\rm exp}\left[ - \sum_{rest} \; |r_m-r_n|^2 \; f_{mn} \right] \sim   1$, the contribution to $P_r$ from the integral $I$ can then be treated almost as a constant.
The above analysis, although a crude approximation, at least provides one information:
the degree of level-repulsion  is reduced 
due to presence  of the localized eigenfunctions
 but is not completely removed. The statistical behavior of the 
ensemble is therefore expected to lie between Poisson and GOE, with 
bulk behavior more close to GOE than that of the edge. This behavior is confirmed by our numerics discussed in \cite{ss}.

For a rigorous analysis, it is necessary to seek alternative routes.  
Fortunately, for applications to real systems, one is interested 
in large $N$-limit of the column constrained matrices;  the statistics 
in this limit can be obtained by another route, based on the mapping 
of eq.(\ref{pe5+}) to another well-known ensemble, namely, Brownian ensemble (BE).  The mapping is based on the analogy of eq.(\ref{pe5a}) to the joint eigenvalue-eigenfunction density of a Brownian ensemble \cite{me, pich},  intermediate between Poisson and Gaussian orthogonal ensemble. As discussed in detail in section V, the analogy can then be applied to seek information about the CCGE spectral statistics  in large $N$-limit.

\vspace{0.2in}

\noindent{\bf Case III:  Independent,  nearest-neighbor Gaussian hopping in $d$-dimension }

\vspace{0.2in}

Consider the case where $H$ represents the Hamiltonian for  a $d$-dimensional lattice of linear size $L$ with a random nearest-neighbor  hopping (Gaussian) and subjected to conditions leading to column constraints (later referred as  CCGE case for arbitrary $d$) . All sites are assumed to be connected by a non-random component too.  Using $N$-dimensional site-basis ($N=L^d$), the ensemble density for the case can again be given by eq.(\ref{rho}) but now

\begin{eqnarray}
\rho_s(H) 
&=&\mathcal{N} \;   {\rm exp}\left[-\gamma \sum_{k=1}^N \sum_{l; l \in z(k)} \; (H_{kl}-b_0)^2 \right] \; \prod_{k, l \not\in z(k)} \delta(H_{kl}-b_1) 
\label{rho3}
\end{eqnarray}
with $b_0, b_1$ as constants and $z$ as the number of nearest neighbors of a site. The symbol $\sum_{l \in z(k)}$ refers to a sum over all nearest-neighbors of a given site $k$ (excluding $l=k$ site) and $\sum_{l \not\in z(k)} \equiv  \sum_{l; l\not=k} - \sum_{l \in z(k)}$.  
As in the case II, here again it is useful to analyze the relative variances of the matrix elements.  
Substituting in eqs.(\ref{cons1}, \ref{cons2}), $\langle H_{kl} \rangle =b_0,  \langle H_{kl} H_{jl}  \rangle = {1\over 2 \gamma} \delta_{kj}$ for a nearest neighbor off-diagonal, and, 
$\langle H_{kl} \rangle=H_{kl}=b_1$, $\langle H_{kl} H_{jl}  \rangle =0$ for a non-nearest neighbor,  the first two moments of the diagonals can be calculated (again with $\alpha_l = \alpha$ for all $l$).  This further gives the ratio of the variance of a typical diagonal to a typical nearest neighbor off-diagonal as
\begin{eqnarray}
\frac{ \langle H_{ll}^2 \rangle - \langle H_{ll} \rangle^2} {\langle H_{kl}^2 \rangle - \langle H_{kl} \rangle^2} = \frac{z-2 \gamma (z b_0+N_z b_1)^2 }{ 1 -  2 \gamma b_0^2}.
 \label{cons4p}
\end{eqnarray}
with $N_z = N-z-1$.
For the case $b_0=b_1=0$ (considered in \cite{ss} for our numerical analysis), this gives  a typical diagonal almost $z$ times of a typical off-diagonal ( therefore weaker than the case II, see eq.(\ref{cons4})). But as  the number of non-zero off-diagonals here  is $z N$, much less than the case II,  a typical eigenfunction is expected to be more  localized than the one in case II.

To develop further insight, it is instructive to rewrite $\delta$-function in eq.(\ref{rho3}) as a limiting Gaussian: $\delta(x-a)  = \lim_{\eta \rightarrow \infty} \; \sqrt{\frac{ \eta}{\pi} }\; {\rm e}^{-  \eta \; (x-a)^2}$. As the modelling of  physical systems by random matrix ensembles is usually  considered in limit $N \rightarrow \infty$, we take $\eta=N$.  Eq.(\ref{rho3}) can now be rewritten as 
\begin{eqnarray}
\rho_s(H)
&=&  C_2 \;  {\rm exp}\left[- \sum_{k=1}^N \left (\gamma \sum_{l; l \in z(k)} \; (H_{kl}-b_0)^2
+ N \sum_{l; l \not\in z(k)} \; (H_{kl}-b_1)^2 \right)\right]  
\label{rho3p}
\end{eqnarray}
with $C_2 = \mathcal{N} \; \left(\frac{2 N}{\pi} \right)^{N N_z/2}$.

To derive the eigenvalue density in this case, we proceed as follows.
Substitution of relation $\sum_{l \not\in z(k)}  H^p_{kl} =\sum_{l; l\not=k}  H^p_{kl} -\sum_{l \in z(k)}  H^p_{kl}$ in eq.(\ref{rho3p}) reduces it as 
\begin{eqnarray}
\rho_s(H) = C_3 \; \rho_{s1} . \rho_{s2} 
\label{rho3g}
\end{eqnarray}
with 
\begin{eqnarray}
\rho_{s1} 
= {\rm exp}\left[- N \sum_{k,l=1; k\not=l}^N \left( H_{kl}^2 -  2 b_1 H_{kl} \right)\right] 
\label{rho3h}
\end{eqnarray}
and 
\begin{eqnarray}
\rho_{s2} 
=  {\rm exp}\left[-  \sum_{k, l; l \in z(k)}^N \;  \left ((\gamma-N) H_{kl}^2 - 2(b_0 \gamma - b_1 N)  \; H_{kl} \right)\right]  
\label{rho3j}
\end{eqnarray}
and $C_3= C_2 \; {\rm e}^{-N (\gamma b_0^2 z + N N_z b_1^2)}$.
Writing $\sum_{k,l; k\not= l} H_{kl} = N \alpha - \sum_{n=1}^N e_n$ and  using eq.(\ref{a3}), $\rho_{s1}$ can be rewritten as 
\begin{eqnarray}
\rho_{s1} 
= {\rm exp}\left[- \frac{N}{2} \sum_{m,n } (e_n-e_m)^2 f_{mn}  +  2 b_1 N^2 \alpha - 2 b_1 N \sum_{n=1}^N e_n \right] 
\label{rho3k}
\end{eqnarray}
 
To express $\rho_{s2}$ as the function of $e_n, O_n$, one can write $H_{kl}=\sum_{m=1}^N e_m O_{km} O_{lm}$ which gives 
\begin{eqnarray}
\sum_{k, l; l \in z(k)} H_{kl}  &=& \sum_{n=1}^N e_n  v_n, \\
\sum_{k, l; l \in z(k)}^N H^2_{kl} &=& \sum_{n,m=1}^N e_n e_m u_{mn} = -\frac{1}{2} \sum_{n,m} (e_n-e_m)^2 u_{mn}, 
\label{suml}
\end{eqnarray}
where 
\begin{eqnarray}
v_{n}  &=& \sum_{k, l \in z(k)}^N O_{kn} O_{ln}, \label{vn} \\
u_{mn} &=& \sum_{k, l; l \in z(k)}^N O_{kn} O_{km} O_{ln} O_{lm}.
\label{umn}
\end{eqnarray}
Here the $2^{nd}$ equality in eq.(\ref{suml})  is due to $\sum_{m=1}^N u_{mn}=\sum_{n=1}^N u_{mn}=0$; the later follows from the orthogonality condition $O^{T} O = I$ which also gives $\sum_{n=1}^N v_n=0$. 
By substituting eqs.(\ref{suml}) in eq.(\ref{rho3j}), followed by a substitution of eqs.(\ref{rho3j}, \ref{rho3k}) in eq.(\ref{rho3g}), 
one can rewrite $\rho_s(H)$  as   

\begin{eqnarray}
\rho_s(H) = C_4 \;  
{\rm exp}\left[-  2 \sum_{n=1}^{N} \omega_n e_n  - \frac{N}{2}\sum_{m, n=1}^{N} (e_m -e_n)^2 \;  (f_{mn}  + (1-\gamma/N) u_{mn} ) \right] 
\label{rho3p}
\end{eqnarray}
with $\omega_n=(N b_1-v_n (b_0 \gamma -b_1 N))$ and $C_4= C_3 \; {\rm e}^{2 N^2 b_1 \alpha}$.

Except for the Gaussian approximation of a $\delta$-function, eq.(\ref{rho3p}) is exact. 
To reduce it further, however, one needs  approximations which can be done as follows. 
As $O_{kN} =\pm\frac{1}{\sqrt{N}}$ (same sign for all $k$, see eq.(\ref{egc3})), it is easy to see from eqs.(\ref{vn}, \ref{umn}) that $v_N = z$ and  $u_{nN}=u_{Nn} =\frac{v_n}{N} ={v_n} f_{nN}$  as $f_{nN}=1/N$.
This  in turn  gives $\sum_{n=1}^{N-1} v_n=-z $ and therefore a typical $v_n \sim -z/(N-1)$.  Further, for $n\not=N$, eq.(\ref{egc3}) gives $O_{ln}=-O_{kn} -\sum_{j\not=l,k} O_{jn}$. This on substitution in eq.(\ref{umn}) reduces it as
\begin{eqnarray}
u_{mn} &=& \sum_{k, l; l \in z(k)}^N O_{kn}^2 O_{km}^2 + \sum_{k, l; l \in z(k)}^N \; \; \sum_{i,j=1;\not=k,l}^N O_{kn} O_{km} O_{in} O_{jm}.
\label{umn1}
\end{eqnarray}
From eq.(\ref{fmn}), the $1^{st}$ term in the above can be rewritten as $ z f_{mn}$. The $2^{nd}$ term, a sum over $4^{th}$ order product of different components of two different eigenvectors,  is expected to be very small (as the components can in general be positive as well negative, resulting in cancellation of terms in the sum). It can therefore be neglected if  $f_{mn}$ is large  which is the case for $O_m, O_n$ pairs at small energy separations $|e_m-e_n|$. Note, similar to discussion in case II given below eq.(\ref{pe5a}), the $O_m, O_n$ pairs with large $|e_m-e_n|$ are more probable to correspond to smaller $f_{mn} \sim 1/N$. The $2^{nd}$ term for such cases can be of the same order as the first term and can not be neglected.   The above suggests that  $u_{mn} \sim  f_{mn}$ for all $m,n$ pairs.
This encourages us to approximate
\begin{eqnarray}
 v_{n} \approx -\kappa, \qquad \qquad u_{mn}+f_{mn} \approx c \; f_{mn}
\label{umn2}
\end{eqnarray}
where $\kappa=z/N$ and $c$ is the ensemble average: $c= \langle \frac{u_{mn}}{ f_{mn} }\rangle $. (Note in general  the ratio $u_{mn}/f_{mn}$ fluctuates for different  $O_m, O_n$ pairs which makes 
eq.(\ref{umn2}) at best a crude approximation. But the results, which it leads to, are supported by our numerical analysis discussed in \cite{ss}). 
Substitution of eq.(\ref{umn2}) in eq.(\ref{rho3p}) reduces it as 
\begin{eqnarray}
\rho_s(H) 
&\approx & C_5 \;  
{\rm exp}\left[-  a_0 \sum_{n=1}^{N-1} (e_n-a_2/a_0)^2  - \frac{c N}{2}\sum_{m, n=1}^{N-1} (e_m -e_n)^2 \;  f_{mn} \right] 
\label{rho3q}
\end{eqnarray}
where $a_0= 1-\kappa(1-\gamma/N) \approx 1-\kappa$, $a_2=\alpha a_0-b_1 N-\kappa(b_0 \gamma-b_1 N) \approx N(1-\kappa) b_1$ and 
$C_5 \approx C_2 \; {\rm e}^{-N^2 N_z b_1^2}$.
The eigenvalue density $P_e(e_1,.., e_N) $ can now be obtained by a substitution of eq.(\ref{rho3q})  in eq.(\ref{pe1}); the transformation $r_n=\; (e_n-a_2/a_0)$ further  reduces it  to 
$\alpha$-free form $P_r(r_1, r_2,.., r_N)$ given by eq.(\ref{pe5}) but now 
\begin{eqnarray}
 P_c (r;O)  &\approx & C_5 \;  
{\rm exp}\left[ - a_0 \sum_{n=1}^{N-1} r_n^2 - \frac{c N}{2}
\sum_{m, n=1}^{N-1} (r_m-r_n)^2  f_{mn} \right]. 
\label{pe9d}
\end{eqnarray}
Note, contrary to eq.(\ref{pe5a}), eq.(\ref{pe9d}) is approximate; its  $\alpha$-independence follows only in the limit $N \rightarrow \infty$ and its dependence on $\gamma$ is not very clear.  An exact formulation for $P_c(r;O)$ in this case could possibly be dependent on the ratio $\alpha/\gamma^{1/2}$.

As $c/a_0 >1$, a  finite basis-connectivity in CCGE seems to reduce the degree of repulsion among its eigenvalues (see discussion below eq.(\ref{pe5a})). The latter being a signature of the eigenfunction-localization, this effect is similar to the unconstrained ensembles with finite basis-connectivity.

\vspace{0.2in}

\noindent{\bf Case IV:  Independent off-diagonals with bimodal distribution }

\vspace{0.2in}

Let us now consider the case with all off-diagonals bimodal distributed (later referred as CCBE case for $d=\infty$). The ensemble density can again be given by eq.(\ref{rho})  but now 
\begin{eqnarray}
\rho_s(H) =\left( \prod_{k,l;  k < l}^{N} \left[ \delta(H_{kl} -{a}) + \delta(H_{kl} + {a} ) \right] \right)
\label{birho1}.
\end{eqnarray}

Using  the representation of a delta function as a limiting Gaussian in large size limit i.e $\delta(x-a)  = \lim_{N \rightarrow \infty} \; \sqrt{\frac{2N}{\pi} }\; {\rm e}^{- 2N\; (x-a)^2}$, eq.(\ref{birho}) can be written as
\begin{eqnarray}
\rho_s(H) = \lim_{N \rightarrow \infty} \; \; \left(\frac{2N}{\pi} \right)^{N(N-1)/2}  \; \sum_{p} 
{\rm exp}\left[- N \; \sum_{k,l; k \not= l} (H_{kl} - b_{p;kl})^2 \right]
\label{birho2}
\end{eqnarray}
where $\sum_p$ refers to sum over all possible combinations of $M = N(N-1)/2$ variables $b_{p;kl}$, with each taking one of the two possible values: $b_{p;kl} = \pm a$ for $k\not=l$. Note eq.(\ref{birho2}) expresses the bimodal ensemble density as  a sum over Gaussian ensemble densities which can be referred as its "Gaussian components". 
Using $H_{kl}=\sum_{n=1}^N e_n O_{kn} O_{ln}$, one can write  
\begin{eqnarray}
\sum_{k,l;k\not=l} b_{p;kl} H_{kl} =  \sum_{n=1}^N e_n \; g_{pn}
\label{sumk}
\end{eqnarray}
 where 
\begin{eqnarray} 
g_{pn} =  \sum_{k,l; k\not=l} b_{p;kl} \; O_{kn} \; O_{ln}. 
\label{gn}
\end{eqnarray}
The orthogonal nature of $O$ gives $\sum_{n=1}^N \; g_{pn} =0$. 
Using eq.(\ref{a3}) and eq.(\ref{sumk}) along with the relation  $\sum_{k,l;k\not=l} b_{p;kl}^2 =N(N-1) a^2$, eq.(\ref{birho2}) can 
be rewritten as 
\begin{eqnarray}
\rho_{s} (E;O) =  \lim_{N \rightarrow \infty} \; \;  \left(\frac{2N}{\pi}  \right)^{N(N-1)/2}  \;
{\rm exp}\left[- \frac{N}{2}  \; \sum_{m,n=1}^{N} |e_m-e_n|^2 \; f_{mn}\right]  \; B(e;O)
\label{pe11a}
\end{eqnarray}
with
\begin{eqnarray}
B(e;O) 
&=&   \sum_{p} {\rm exp}\left[-(N-1)N^2 a^2 +  2 N \sum_{n=1}^N \; g_{pn}\; e_n \right].  
\label{B0}
\end{eqnarray} 

The joint probability density $P_{e}(e_1,..e_N)$ for this case can now be given by  eq.(\ref{pe1}) with $\rho_s$ as in eq.(\ref{pe11a}). As clear from a comparison of eqs.(\ref{pe4}, \ref{pe11a}), $\rho_s$ for this case is same as that of case II except for the term $B(e;O)$ and $\gamma=N$. 
Using $r_n=\sqrt{N} (e_n-\alpha)$, and, following similar steps as used in the derivation of eq.(\ref{pe5+}) from eq.(\ref{pe4}), the density $P_{r}(r_1,..r_N)$ for eq.(\ref{birho2}) now becomes 
\begin{eqnarray}
P_{r} (r_1, r_2,.., r_N) 
&=&   \int \; P_c (r;O) \; B(r;O) \; \delta_o(O) \;  {\rm D}O 
\label{pe12}
\end{eqnarray}
with $P_c(r;O)$ given by eq.(\ref{pe5a}) and 
\begin{eqnarray}
B(r;O) 
&=& \lim_{ N \rightarrow \infty} \;\; \sum_{p} {\rm exp}\left[-(N-1)N^2 a^2 + 2\sqrt{N} \sum_{n=1}^N \; g_{pn}\; r_n \right]  
\label{B}
\end{eqnarray}
Clearly only those terms of the $\sum_p$ contribute  to $B(r;O)$ for which 
the exponent vanishes i.e $2 \sum_{n=1}^N g_{pn} \; r_n = \sqrt{N} \; N(N-1) a^2$ or equivalently, from eq.(\ref{sumk}), 
\begin{eqnarray}
2 \sum_{k,l; k\not=l} b_{p;kl} H_{kl} =  N(N-1) a^2.
\label{ge}
\end{eqnarray}
As left side of the above condition contains $N(N-1)$ terms, with each $H_{kl}$ Gaussian distributed with mean $b_{p;kl}$, it can be satisfied in many ways e.g. if $H_{kl} = b_{p;kl}/2= \pm a/2$ for all $k,l$ pairs  or $H_{kl}= b_{p;kl}$ for $3 N(N-1)/4$ pairs and $H_{kl}=- b_{p;kl}$  for rest $N(N-1)/4$ of them. Thus the eigenvalue probability density for bimodal case is dominated, in large $N$-limit,  by those Gaussian  components  in which almost all $H_{kl}$ are of the same order  as their mean $b_{p;kl}$.

The above gives $B(r;O)$ as a constant (equal to number of terms in $\sum_p$ satisfying the condition $(\ref{ge})$).   This alongwith $P_c(r;O)$ given by eq.(\ref{pe5a}), $P_r$ in the bimodal case reduces to a same form as the Gaussian case II. This is also reconfirmed by the numerically observed  analogy  of the spectral fluctuations for both the cases (see  \cite{ss}).

\vspace{0.2in}

\noindent{\bf Case V:  Independent, nearest-neighbor bimodal hopping in $d$-dimension }

\vspace{0.2in}

As in the case III, let us again consider the dynamics in a $d$-dimensional  lattice of linear size $L$ but  random component of the nearest-neighbor  hopping is now chosen to be bimodal type. All sites are assumed to be connected by a non-random component too. In the $N$-dimensional site-basis ($N=L^d$), the ensemble density for the case is given by eq.(\ref{rho})  ((later referred as CCBE case for arbitrary $d$) but now 
$\rho_s$ is as follows:
\begin{eqnarray}
\rho_s(H) = \mathcal{N} \; \left(\prod_{k,l;  l \in z(k) }^{N} \left[ \sum_{q=\pm 1} \delta(H_{kl} - q a ) \right] \right) \; 
\left(\prod_{k,l; l \not\in z(k)}^N \delta \left( H_{kl} - b_1 \right) \right) 
\label{birho6}.
\end{eqnarray}
Proceeding exactly as in the case IV above, the eigenvalue density $P_r(r_1,..r_N)$ in this case can be shown to be analogous to that of the 
 nearest neighbor Gaussian hopping case III i.e eq.(\ref{pe9d}).

\section{ Spectral fluctuations of column constrained ensembles}

 In past, there have been several attempts to study the spectral statistics of  column/row constrained 
matrices (see for example (\cite{mm, yvf, os, bray0} and references there in); the presence of correlations 
among  their  columns/rows  makes the determination a technically non-trivial  task. Previous studies \cite{mm, luck, bray0, os, yvf}, often using field theoretic approach, have analyzed the 
$1-$ and $2$-point correlations of the level-density (in the spectral-bulk) for infinite range CCEs. Here we consider a different approach, based on a mapping of the joint probability density of 
the eigenvalues of a CCE  to that of a Brownian ensemble (BE) (of real-symmetric matrices, without 
column constraint). The BEs  have been 
extensively studied during previous decade and a great deal of analytical/numerical information about 
their statistical fluctuations  is available \cite{dy, rp, me, ap, fkpt, alt, fgm, ls, ksh, pich, shapiro, 
ks, forres, psc, gp, hb}. The mapping facilitates the  available information directly to be applied to an
infinite range CCE which in turn helps in improvement of the previous results for $2^{nd}$ order correlations 
as well as determination of the higher orders. 
The approach also helps us to analyze the CCEs representing more generalized system 
conditions, e.g CCE with nearest neighbor hopping etc with both Gaussian and bimodal disorder. 
The mapping is relevant for another reason too. As discussed in \cite{psand, ps-all}, the 
spectral fluctuations of the generalized random matrix ensembles without column constraints can be 
expressed in terms of  the BEs. The present study therefore connects an ensemble with column constraints 
to many other ensembles without them. 

Before discussing the details of the mapping, we  briefly review the Brownian ensemble first:

\subsection{Brownian ensembles (BE): relation with column constrained ensembles}

A Brownian ensemble of Hermitian matrices $H$ can in general be described as a non-stationary 
state of the matrix elements undergoing a cross-over due to a random perturbation of a stationary ensemble, say $H_0$,  by another one, say $V$:  $H=\sqrt{f} (H_0+\lambda V)$ with $f=(1+\lambda^2)^{-1}$ (\cite{me, ap}). The type of a BE, appearing during the 
cross-over, depends on the nature of stationary ensembles $H_0, V$ and their different pairs may 
give rise to different BEs \cite{ap, psijmp}. The present knowledge of ten types of stationary 
ensembles \cite{zirn} leads to possibility of many such cross-overs and, consequently, 
many types of BEs.

In context of the column constrained ensembles of real-symmetric matrices, the relevant BE is the one  appearing during a transition from Poisson $\rightarrow$ Gaussian orthogonal  ensemble (GOE) \cite{psand}. With $H_0$, $V$ as $N \times N$ matrices taken from  Poisson and GOE respectively, the BE in this case is an ensemble of  real-symmetric matrices $H$, free from any column constraint and  described by the probability density
\begin{eqnarray}
\rho(H)  \propto 
{\rm exp}{\left[-\frac{\eta}{2} \; \sum_{i=1}^{N} H_{ii}^2 -  
\frac{\eta}{2} (1+\mu) \sum_{i,j=1; i \not= j}^{N} H_{ij}^2 \right]} 
\label{be1}
\end{eqnarray}
with $\eta$ as an arbitrary parameter and $(1+\mu)=(\lambda^2 f)^{-1}$; here $H=H_0$ for $\lambda \rightarrow 0$ or 
$\mu \rightarrow \infty$ and $H \rightarrow V$ for $\lambda \rightarrow \infty$. 
An ensemble  of $H$ matrices given by the above measure, is also known 
as Rosenzweig-Porter (RP) ensemble \cite{rp}. 
As $H$ is a real-symmetric matrix, the jacobian of transformation from $H$-space to eigenvalue-eigenfunction-space is $J(r, O|H) \propto \Delta_{N}(r)$ (eq.(\ref{Delta})), with $r_n$ as  the eigenvalue  and $O_n$  corresponding eigenfunction (subjected only to orthonormality constraint) for $n=1,2,\ldots N$ \cite{me}. 
The joint eigenvalue-eigenfunction density $P_b(r, O)$ can again be obtained from the relation  $P_b(r,O) \; {\rm D}r \; {\rm D}O = \rho(H) \; {\rm D}H $, with ${\rm D}H= J(r, O|H)  \; {\rm D}r \; {\rm D}O$, which gives 
\begin{eqnarray}
P_b(r; O) &=&   \Delta_{N}(r) \; \rho \left(O^{\dagger} \; r \; O \right) \nonumber \\
&=&  \mathcal{N} \; \Delta_{N}(r)  \;
{\rm exp}\left[-\frac{\eta}{2} \;  \sum_{m=1}^{N} r_m^2 -
\frac{\eta \; \mu}{4} \; \sum_{m, n=1; m \not=n}^{N} |r_m-r_n|^2 \; f_{mn} \right]
\label{peb} 
\end{eqnarray}
where $\mathcal{N}$ is the normalization constant and $f_{mn}$ is same as in eq.(\ref{fmn}).
An integration of eq.({\ref{peb}) over $O$-space  (subjected only to orthogonality constraint) gives the joint eigenvalue density $P_r$ of BE: $P_{rb}(r_1, r_2,.., r_N) =   
  \int \; P_b (r; O) \; \delta(O^{\dagger}O-I) \;  {\rm D}O $ which can further be used to calculate the $n^{th}$ order spectral correlation $R_n(r_1,\ldots, r_n)$,  the probability density of finding $n$ eigenvalues at $r_1, r_2,..,r_n$,   defined as   
\begin{eqnarray}
R_{n}(r_1,\ldots, r_n) &=&
\langle \sum_{i_1,\ldots, i_n=1}^N \delta(r_1-e_{i_1}) \ldots \delta(r_n-e_{i_n}) \rangle  \label{rr02} \\
&=& \frac{N !}{(N-n)!} \; \int P_{r}(r_1,\ldots, r_N) \; \prod_{k=n+1}^{N} {\rm d}r_k. 
\label{r02}
\end{eqnarray} 
Using the above definitions,  $R_n$ for BE (later referred as $R_{nb}$) can be written in terms of  $P_b(r;O)$: $R_{nb}(r_1,\ldots, r_n) =\frac{N !}{(N-n)!} \; \int \; P_b (r; O) \; \delta(O^{\dagger}O-I) \;  
 \prod_{k=n+1}^{N} {\rm d}r_k \; {\rm D}O$.

All spectral fluctuation measures can in principle be obtained from  $R_n$ \cite{me}. As eq.(\ref{peb}) indicates, the  BE is a  basis-dependent ensemble with statistics governed  by a single parameter $\mu$ and thus appears, although deceptively, easier to analyze. In past few decades, there have been several attempts to derive an analytical formulation of their fluctuations but very few exact results are known. The reason lies in  unavailability of the results for the integration over orthogonal  matrix space (although some progress has been made during past decade, see for example \cite{itz, gor}). Note the unitary space integration, which allows the determination of BE statistics during Poisson to GUE transition \cite{alt, fgm, gp, hb}, has already been achieved \cite{me}.   

In an attempt to circumvent the above difficulty, a new approach has been introduced during last decade \cite{psand, ps-all, psijmp}. Based on the single parametric formulation of the diffusion of their probability density, 
the approach connects the BEs to a wide range of multi-parametric Gaussian ensembles of the same global constraint class. The mapping of  a column constrained ensemble to the BE therefore  connects the former to many other ensembles (e.g Anderson ensemble) which appear in different areas and have been studied by area-specific tools \cite{psand}. The information from these studies can then be applied to 
probe the column constrained ensembles.

\vspace{0.2in}

{\bf Comparison of CCE cases II, IV  with BE:} For clarity, let us first compare the CCGE case (II) of section IV with BE.  As clear from eq.(\ref{peb}),  the probability $P_b(r;O)$ that one of the BE-eigenvalues, say $r_N=0$, and corresponding eigenfunction $O_N$ is extended i.e $O_{kN} =  \frac{1}{\sqrt{N}}$, is of the  same form as $P_{c}(r;O)$ given by eq.(\ref{pe5a}) with $\eta=\frac{2}{N}, \mu=N$:

\begin{eqnarray}
P_{c}(r; O) &=& P_b (r; O) \; \delta(r_N) \; \delta\left(O_{N}-\Phi\right) 
\label{peb3}
\end{eqnarray}
where $\Phi$ is a column vector: $\phi_{k}=\frac{1}{\sqrt{N}}$ for $k=1,\ldots N$. (Note for $O_N =\Phi$, the orthogonality condition $\sum_{k=1}^N O_{kn} O_{k N} =\delta_{nN}$ ensures that $\sum_{k=1}^N O_{k n} =0$ for all other eigenfunctions with $n \not=N$). This permits us to express the  spectral correlations of  CCE in terms of those of the BE as follows. 
The definition (\ref{r02}) of $R_n$ along with eq.(\ref{pe5}) gives  
 $R_{n}$ for CCE, referred as $R_{nc}$, 
\begin{eqnarray}
R_{nc}(r_1,\ldots, r_n)  &=& \frac{N !}{(N-n)!} \; \int P_{c}(r;O) \; \delta_O(O) \;  \prod_{k=n+1}^{N} {\rm d}r_k \; {\rm D}O 
\label{rnc}
\end{eqnarray}
with $\delta_O(O)$  defined in eq.(\ref{delo}). But note the latter is just the  orthogonality condition along with a non-random, uniform eigenfunction: $\delta_O(O) = \delta(O_N-\Phi) \delta(O^T O-I)$. 
Substitution of eq.(\ref{peb3}) relates  $R_{nc}$ to BE statistics  
\begin{eqnarray}
R_{nc}(r_1,\ldots, r_n)  
&=& \frac{N !}{(N-n)!} \;  \int P_{b; N-1}(r;O) \; \delta(O^T O-I) \; \prod_{k=n+1}^{N-1} {\rm d}r_k \; {\rm D}O  
\label{peb6}
\end{eqnarray}
where $P_{b,N-1}(r;O) \equiv P_{b}(r_1,\ldots, r_{N-1}, 0;O_1,\ldots,O_{N-1}, \Phi)$ 
refers to the section  of joint eigenvalue-eigenfunction density $P_b(r;O)$  for a $N \times N$  BE on the $r_N=0, O_N=\Phi$ plane. This can equivalently be viewed as a sub-ensemble of the BE, consisting of the matrices each one of which has a zero eigenvalue and corresponding eigenvector as uniformly extended. 
Using the definition of $R_n$ (eq.(\ref{r02})),  the right side of eq.(\ref{peb6}) can be written as  the $n^{th}$ order correlations ${\tilde R}_{nb}(r_1,\ldots, r_n)$ of the BE sub-ensemble     
\begin{eqnarray}
R_{nc}(r_1,\ldots, r_n)  = {\tilde R}_{nb}(r_1,\ldots, r_n) 
\label{rnbc}
\end{eqnarray}
where ${\tilde R}_{nb}(r_1,\ldots, r_n) =\frac{N !}{(N-n)!} \; 
\int \; P_{b;N-1} (r; O) \; \delta(O^{\dagger}O-I) \;  
\prod_{k=n+1}^{N-1} {\rm d}r_k \; {\rm D}O$.

Using eq.(\ref{rr02}), it is easy to express the correlations in the sub-ensemble  in terms of those of the BE (see Appendix C)
%. 
\begin{eqnarray}
{\tilde R}_{nb} (r_1,\ldots, r_n)  \approx  R_{nb} (r_1,\ldots, r_n) \; \prod_{k=1}^n \left(1+\delta(r_k)\right) 
\label{rnbc1}
\end{eqnarray}
Applying eqs.(\ref{rnbc}, \ref{rnbc1}) for $n=1$, the ensemble averaged level-density  $R_1(r)$ of the CCE  can be expressed in terms of that of the BE: 
\begin{eqnarray}
R_{1c}(r)  = R_{1b}(r) \left(1 + \delta(r)\right).
\label{rnbc2}
\end{eqnarray} 
Thus the ensemble averaged level density of the CCE is expected to deviate from that of the BE near zero energy; (this is also confirmed by the numerics shown in figure 12(a) of \cite{ss}).
Similarly, using eqs.(\ref{rnbc}, \ref{rnbc1}) for $n>1$,  one can write higher order spectral correlations of the CCE in terms of those of a BE: 
$R_{nc}(r_1,\ldots, r_n)  =  R_{nb}(r_1,\ldots, r_n)$ for $r_1,\ldots,r_n \not=0$.

Alternatively, the CCGE case II can also be mapped to a $(N-1) \times (N-1)$ BE as follows: except for the constraint $\delta(r_N)$, eq.(\ref{pe5a}) is analogous to  eq.(\ref{peb}) with $\eta=\frac{2}{N}, \mu=N$. Although the $O$-space integration in the CCGE case is subjected to constraints  given by $\delta_O$, the latter just reduces the size of the $O$-matrix space from $N$ to $N-1$.  The number of independent eigenfunction components  in a $N\times N$ column constrained matrix is therefore same as that for a $(N-1) \times (N-1)$ real-symmetric Brownian matrix. An integration over eigenfunction-space of eq.(\ref{pe5}) then renders an eigenvalue distribution for a CCGE analogous to that of a BE, except for a logarithmic potential term $\sum_k \log |r_k|$ in the exponent; the term originates from the level-repulsion of all other eigenvalues from the non-random one. 
As each term of type $\log |r_k|$ in the exponent of eq.(\ref{pe5a}) appears along with a quadratic term of type $r_k^2$, the former is negligible as compared to later for large $r_k \ge 1$. Furthermore 
as the level-repulsion between all non-zero eigenvalue pairs is same for both CCGE and BE, the local correlations in the $(N-1)$ dimensional eigenvalue space (excluding zero eigenvalue) of a $N \times N$ CCGE are  expected to be analogous to that of a $(N-1)\times (N-1)$ BE. The analogy is confirmed by the numerics given in \cite{ss}. 

Due to its eigenvalue-eigenfunction density being analogous to that of the CCGE case II, the above discussion is also applicable to CCBE case IV.

 As already mentioned,  there have been a number of attempts in past to obtain the spectral statistics of CCEs e.g the level density for random impedance networks in \cite{luck} and random master equations in \cite{bray0}, the level density and 2-point spectral (bulk)   correlation $R_2(r)$ (with $r=|r_1-r_2|$) of Euclidean matrices in \cite{mm} and  \cite{os} respectively (a field theoretic formulation), $R_2(r)$ for random reactance network (again for bimodal distribution, infinite range and only in spectral-bulk). Reducing $R_2(r)$ to an integral form similar to that of a GOE, the studies \cite{os, yvf} conclude that $R_2(r)$ in the spectral bulk of an infinite range CCE is analogous to that of a GOE; these studies are based on the field-theoretic approach using saddle point approximation. Our analysis however indicates that $R_2(r)$ of CCE is analogous to that of a BE (eq.(\ref{be1}) with $\mu=N$) which approaches a GOE near zero energy. But the  BE-statistics  deviates from  that of a GOE, significantly at large spectral-ranges (\cite{pich}, see also discussion below eq.(\ref{alm}) ). Thus our $R_2(r)$-result for CCE does not essentially contradicts the result of \cite{yvf} but provides an improved version applicable for ranges beyond the spectrum center.

\vspace{0.2in}

{\bf Comparison of CCE cases III, V with BE:}  We now compare eq.(\ref{pe9d}) for the case III of section IV with eq.(\ref{peb}).  As in the case II,   the density $P_c(r;O)$  for this case can again be related to a BE given by eq.(\ref{peb3}), with 
$P_b(r;O)$ given by eq.(\ref{be1}), but now $\mu = (c/a_0) N$ with $c/a_0>1$. Eq.(\ref{rnbc}) is therefore valid for this case too (with ${\tilde R}_{nb}$ now representing the $n^{th}$ order correlation of the corresponding BE analog). 

As suggested by the crude approximations,  the parameter $\mu$ for the BE analog in this case seems to be larger than the $d=\infty$ CCGE case. An increased $\mu$ corresponds to an eigenvalue statistics  shifted more 
towards Poisson distribution (see \cite{psand, pich}); this is indeed supported by  our numerical analysis \cite{ss}.

\subsection{Fluctuations measures of CCE: exploiting BE connection}

Eq.(\ref{peb})  indicates the non-stationary nature of the BE statistics: it varies significantly from middle of the band to the edge;  (numerically confirmed in \cite{ss}). As shown by a previous analysis \cite{pich} of the BE with finite $\mu$, the levels with sufficiently close energies i.e for energy intervals smaller than $E_{\mu}=1/\sqrt{\mu}$ repel each other like a GOE. 
This can also be explained by an alternative formulation for the BE-statistics:    the spectral fluctuations around $R_1(r)$  are 
governed by a parameter which is the typical off-diagonal square measured in units 
of the local mean-level spacing $\Delta_{local}(r)$ (referred as the 
spectral complexity parameter) \cite{psand, me}): 
\begin{eqnarray}
\Lambda(r)= \frac{1}{(1+\mu) \Delta_{local}^2(r)}
\label{alm}
\end{eqnarray}
Here $\Delta_{local}(r)$ is the local mean-level spacing: $\Delta_{local}(r) = (R_1(r))^{-1}$ for a BE. 
A variation of $\mu$ changes  the level-density  $R_1(r)$ of the BE from a Gaussian ($\mu \rightarrow \infty$) to a semi-circle ($\mu \rightarrow 0$)  \cite{shapiro}. This changes $\Lambda(r)$ and therefore results in a cross-over of the BE statistics from Poisson ($\Lambda \rightarrow 0$) to GOE ($\Lambda \rightarrow \infty$) at a fixed $r$.  As discussed in \cite{shapiro}, for $\mu=\nu N$, $R_1(r) = N \; F(r)$  with $F(r) \sim \frac{1}{\sqrt{\pi}} \; {\rm e}^{-r^2} $ for $\nu \gg 1$. For  $\nu \ll 1$,  $F(r) \approx \frac{1}{\pi} \; \sqrt{2 \nu - \nu^2 r^2}$ for $ r^2 \ll 1/\nu$ but develops Gaussian tails for large $|r|$.  Although the results for $\nu=1$ are not known analytically, 
numerical studies indicates a semicircle behavior \cite{shapiro} in the bulk {: $F(r) =\frac{1}{\pi N a} \sqrt{2a N-r^2}$ with $a$ as a constant (see figure 12(a) in \cite{ss}).  As  the mean level density in the middle of the band is relatively higher than the edge, the BE-statistics for a fixed $\mu$ tends to GOE in the bulk but remains Poisson near the edge.

Eq.(\ref{alm}) indicates that, for the cases $\mu=\nu N$ with $\nu \sim 1$,  $\Lambda(r)$ becomes size-independence in the bulk regime. As the level-statistics is governed by $\Lambda$ only, it is size-invariant as well as intermediate between Poisson and GOE even in the limit  $N \rightarrow \infty$ and is therefore termed as critical. 
These critical BEs form a one parameter family of non-equilibrium ensembles lying between Poisson and GOE equilibrium, and, with a size-independent level-statistics. Following analogy with BE, the level-statistics of the CCE is expected to   approach an invariant form, intermediate between Poisson and GOE, in large $N$-limit; this is indeed  confirmed by our numerical analysis \cite{ss}.

Using the CCE-BE mapping,  the BE results with $\mu=c N$ can  directly be used for $N \times N$ column-constrained Gaussian and bimodal cases described by eqs.(\ref{rho2},\ref{rho3}, \ref{birho1}, \ref{birho6}); some of these results are briefly reviewed in appendix B (for more details see \cite{ap}).

\section{Conclusion}

In the end, we summarize with our main results and open questions.

Here we have analyzed the role of a specific global constraint for complex systems,  appearing as column/row sum rules on their matrix representations and in combination with other local and global constraints. 
Our study, focussed here on  column constrained real-symmetric matrices,  reveals some of their important features e.g a non-random eigenvalue 
with corresponding eigenvector uniformly extended, the lack of single basis-state localization for a typical eigenvector
which is in constrast with an unconstrained real-symmetric matrix. 
We also  find  that, in large size limit,  the column constrained matrices are statistically analogous to a special type of critical BE, intermediate between Poisson and GOE.  The BE analogs of these matrices are size-independent but depend on the type of their basis-connectivity;  the one for finite connectivity depends on a single parameter  but that for infinite connectivity is free of all parameters. 
The column constrained matrices therefore undergo a cross-over from Poisson to the parameter-free BE with basis-connectivity  as the transition parameter. The exact value of the column constant  has no non-trivial influence either on the matrix properties or the ensemble ones if it is 
same for all columns and the matrix is real-symmetric; this however may not be valid for more general cases e.g varying column constants, complex matrices etc.  

As well-known, a real-symmetric ensemble without column-sum rule (a system with no other constraints except time-reversal symmetry) undergoes a Poisson to GOE transition but the column sum rule inhibits the cross-over from reaching to  GOE.  The inability of column-constrained ensemble to reach GOE at typical energy-scales (other than zero energy) can be explained as follows: the correlations arising from the column sum-rule  make the diagonals  effectively  much larger than the off-diagonals (their ratio  dependent on the number of non-zero elements in a column). This tends to localize the dynamics  around the basis-states which however is opposed by the basis-connectivity (hopping etc). In absence of the column sum-rule, the spectral statistics is governed by a competition between the diagonal disorder and hopping, resulting in a crossover from Poisson to GOE universality class but, in its presence, the  disorder always dominates causing the equilibrium to occur  midway between Poisson and GOE.  The statistics in this equilibrium represents a new universality class, free of all parameters, and, obtained by  imposing an additional symmetry (e.g. Goldstone symmetry) leading to column constraint  along with time-reversal symmetry. Note this  class is different from the ten well-known standard universality classes \cite{zirn} which correspond to the infra-red  renormalization group fixed points describing the ergodic limit). This is because, contrary to  the previously known classes,  the statistics in a column constrained ensemble approaches a fixed point lying in the non-ergodic regime \cite{psr}.   (Note the universality of the spectral fluctuations of a column constrained ensemble follows from its mapping to a specific BE with an  ensemble  density dependent only on a single parameter namely matrix size $N$. As discussed in section V.B, the level-statistics of this BE is size-independent, free of any parameters and  is universal in the sense that all systems modeled by this BE will have same statistics at a given energy. However  the eigenfunctions of the BEs appearing between Poisson and GOE universality classes are known to be partially localized which manifest in non-ergodicity of the spectral fluctuations \cite{pich}. The CCE-BE mapping further implies a similar universality and non-ergodicity of the statistics  for all systems modeled by CCEs, e.g. the systems discussed in section II. Appearance of nine more universality classes of CCEs is expected on similar grounds, arising in presence of the following combinations: Goldstone symmetry and no time-reversal (leading to complex-Hermitian ensembles with column constraints), Goldstone symmetry and half-integer angular momentum  (leading to real-quaternion ensembles with column constraints), Goldstone symmetry along with one of three chiral symmetries (chiral ensembles with column constraints),  Goldstone symmetry along with one of the four particle-hole symmetries (particle -hole ensembles with column constraints).

It is worth noting  that a BE   itself is a non-equilibrium state of a disorder-driven transition in space of ensembles subjected to a single global constraint i.e  time-reversal symmetry. The appearance of an almost BE type ensemble as an equilibrium state of the transition in column constrained matrices  (two global constraints here, namely, time-reversal and column constraint) therefore suggests a hierarchy of equilibriums: 
the non-equilibrium states of the transition in the ensemble-space with  lesser number of global constraints appear as the equilibrium states of the transition in the space with higher constraints.

 As revealed by several studies in past, complexity gives birth to a great deal of hidden connections in seemingly very different scientific areas.
Seeking the connections (i.e universality) is one of the most beautiful as well as useful aspects of scientific research.} We emphasize that an important feature of our present analysis is revealing the connection between column constrained matrices and Brownian ensembles. Connection among  ensembles within same global constraint class has been reported in past; for example, the statistics of the BEs and disorder Hamiltonians e.g. Anderson Hamiltonian, both with time-reversal symmetry, are analogous if their complexity parameters are same \cite{psand}. The present work further extends it by connecting the ensemble with different global constraint classes.  As discussed in previous section, the CCE-BE connection also helps to improve the results obtained by previous studies \cite{os, yvf}. Another important connection not discussed here but worth exploring is between sparse or banded CCEs and the ensembles with enhanced diagonals (different from the BE or Rosenzweig-Porter ensemble); the latter  appear in many interesting physical contexts e.g. two particle localization \cite{shep3}.

In recent years, there have been a lot of interest in statistical analysis of 
the bosonic excitations. The study presented here reveals the existence  of a new universality class in the spectral statistics of Goldstone modes but it does not  provide information about the eigenfunctions. The latter requires a study of  the ensembles with both column constraints as well as an additional constraint (leading to an extended mode at the minimum eigenvalue).  Similarly, for other applications, an investigation of CCE with different symmetry constraints and conservations laws is needed (e.g.  the non-Hermitian ensembles with column constraints are applicable in Google matrix analysis).

Previous studies of the system-dependent random matrix ensembles with  anti-unitary symmetries as the global constraints indicate a single parametric dependence of the fluctuation measures \cite{psijmp, psand, ps-all}.  The search for a  similar formulation  for  column-constrained ensembles  is 
desirable too; it will provide a common theoretical formulation for the cases with lower or higher basis-connectivity, anisotropic hopping, correlated off-diagonals etc. Our attempts so far in this direction are encouraging.

\section*{Acknowledgment}

We thank Robin Stinchcombe for educating us about the column sum rule in systems with Goldstone modes  and many helpful discussions at an  initial stage of this project. We also gratefully acknowledge helpful discussions with Michael Berry, Deepak Kumar and  John Chalker.

\appendix

\section{Derivation of eq.(\ref{pe4})}

Consider a real-symmetric $N \times N$ matrix $H$ with an eigenvalue $e_n$ and corresponding eigenfunction $O_n$, with its components referred as $O_{kn}$ for 
$n=1 \rightarrow N$. Using the relation $H_{kl}= \sum_{n=1}^N e_n O_{kn} O_{ln}$,  
a sum $S_1$ over  all upper (or lower) off-diagonal matrix element squares of $H$ 
can be reduced in following form: 
\begin{eqnarray}
S_1 &=&  \sum_{k,l=1; k \not= l}^{N} H_{kl}^2 \\
&=& \sum_{k,l=1; k \not= l}^{N} \sum_{m,n=1} e_n e_m O_{kn} O_{ln} O_{km} O_{lm} \\
&=&  \sum_{m,n=1}^N e_n e_m \left[ \sum_{k,l} \left(O_{km} O_{kn} \right) 
\left(O_{lm} O_{ln} \right) -  \sum_{k=1}^N O_{kn}^2 O_{km}^2 \right]
\label{a1}
\end{eqnarray}
The orthogonality relation of the eigenfunctions gives $\sum_{k=1}^N O_{kn} O_{km} =\sum_{k=1}^N O_{nk} O_{mk}  = \delta_{nm}$; its substitution in eq.(\ref{a1}) gives  
\begin{eqnarray}
S_1 &=&  \sum_{m,n} e_n e_m \left[\delta_{n m} - 
\sum_{j=1}^N O_{jn}^2 O_{jm}^2 \right] \\
&=& \sum_{n=1}^N e_n^2 + \sum_{m,n=1}^N e_n e_m 
\left(\sum_{j=1}^N O_{jn}^2 O_{jm}^2 \right) \\
&=& \sum_{n,j=1}^N e_n^2 O_{jn}^2  - \sum_{j=1}^N \sum_{m,n=1}^N e_n e_m  O_{jn}^2 O_{jm}^2 \\
&=& {1\over 2} \; \sum_{j=1}^N \sum_{m,n=1}^N \left(e_n^2 + e_m^2  - 2 e_n e_m \right)  O_{jn}^2 O_{jm}^2 \\
&=& {1\over 2} \; \sum_{j=1}^N \sum_{m,n=1}^N \left(e_n-e_m \right)^2  O_{jn}^2 O_{jm}^2 \\
&=&  \sum_{j=1}^N \sum_{m,n=1; m < n }^N \left(e_n-e_m \right)^2  O_{jn}^2 O_{jm}^2
\label{a2}
\end{eqnarray}
The above implies 
\begin{eqnarray}
\sum_{k,l=1; k \not=l}^N H_{kl}^2 =              
{1\over 2} \sum_{j=1}^N \sum_{m,n=1}^N \left(e_n-e_m \right)^2  O_{jn}^2 O_{jm}^2
= \sum_{j=1}^N \sum_{m,n=1; m < n}^N \left(e_n-e_m \right)^2  O_{jn}^2 O_{jm}^2
\label{a3}
\end{eqnarray}
Substitution of eq.(\ref{a3}) in eq.(\ref{rho2}) and subsequent use of eq.(\ref{pe1}) gives eq.(\ref{pe4}).      

The above result can also be used to derive eq.(\ref{peb}) from eq.(\ref{be1}). By rearranging the terms and using the real-symmetric nature of $H$, eq.(\ref{be1}) can be rewritten as 
$\rho(H) \propto {\rm exp}[- S_2]$ where $S_2$ is given as  
\begin{eqnarray}
S_2 &=& {\eta \over 2} \sum_{i=1}^{N-1} H_{ii}^2 +  \eta (1+ \mu) \sum_{k,l=1; k < l}^{N-1} H_{kl}^2 \nonumber \\
& = &  {\eta \over 2} \left( \sum_{k,l=1}^{N-1} H_{kl}^2 + \mu \sum_{k,l=1; k\not=l}^{N-1} H_{kl}^2 \right).
\label{a4}
\end{eqnarray}
As the first term of eq.(\ref{a4}) on RHS is the trace of the square of a $(N-1) \times (N-1)$ matrix $H$, we have $$\sum_{k,l=1}^{N-1} H_{kl}^2 = \sum_{n=1}^{N-1} e_n^2 $$. This along with relation (\ref{a3}) (now applying it for a  matrix $H$ of size $N-1$) leads to eq.(\ref{peb}).     

%Following similar steps as above, eq.(\ref{pe9}) can also be obtained.

\section{Some spectral fluctuation measures for Brownian ensemble with $\mu=cN$}
 
Here we briefly review some of the spectral fluctuation measures of the BE which, using CCE-BE analogy, can directly be applied for a CCE:

%\noindent{\it Level-density $R_1(r)$}:  In the spectrum-bulk, $R_1(r)$  is a semi-circle in large $N$-limit: $R_1(r) =\frac{1}{a \pi} \sqrt{2a N -r^2}$ with $a$ as a constant. An exact numerical  diagonalization  of eq.(\ref{rho2}) given in \cite{ss} confirms  $R_1$-behavior of the ensemble described by eq.(\ref{rho2}). This result is also in agreement  with \cite{luck}. 

%\vspace{0.1in}

\noindent{\it 2-point density correlation $R_2(e_1, e_2)$}: 
Defined as $R_2(e_1, e_2)=\frac{N !}{(N-2)!} \; \int P_e(e) \; \prod_{k=3}^{N-1} {\rm d}e_k$, it gives the probability of finding the  eigenvalues $e_1, e_2$ at a distance $r=|e_1-e_2|$. In large $N$ limit, the small-$r$ behavior of $R_2$ can be described by a closed form equation
\begin{eqnarray} 
{1\over 2} \frac{\partial R_2}{\partial r} = \frac{\partial^2 R_2}{\partial r^2}- \frac{\partial}{\partial r}\left(\frac{R_2}{r}\right)
\label{G1}
\end{eqnarray}
The large $N$, large-$r$ limit behavior for $R_2$, can be given as  (see eq.(23) of \cite{ap})
\begin{eqnarray} 
R_2(r,\Lambda) \approx R_2(r,\infty) +2 \Lambda  \; \int_{-\infty}^{\infty} \frac{ R_2(r-s;\infty) -R_2(r-s;0)} {s^2 + 4\pi^2  \Lambda^2} \; {\rm d}s 
\label{G}
\end{eqnarray}
%where
%$R_2(r,\infty)= {{\rm sin}^2(\pi r)\over \pi^2 r^2} $  and $R_2(r,0)=1$.
%
%\begin{eqnarray} 
%R_2(r,\Lambda) \approx R_2(r,\infty) +\frac{2 \Lambda }{( r^2 + 4\pi^2  \Lambda^2)}
%\label{r2}
%\end{eqnarray} 
where  $R_2(r,\infty)= 1- \left(\frac{\sin \pi r}{\pi r} \right)^2 -\left({{\rm d}\over {\rm d}r} \; \frac{sin\pi r}{\pi r}\right) \; \int_r^\infty {\rm d}x  \; \frac{\sin\pi x}{ \pi x} $ (GOE limit). 

Due to Poissonian nature of unperturbed levels, the multi level-interactions can be neglected within perturbation theory approach. Thus,  
for small $\Lambda$, $R_2$ can be derived from diagonalization of the corresponding $2\times 2$ dimensional submatrix \cite{fkpt}
\begin{eqnarray} 
R_2(r,\Lambda) \approx  \frac{r}{\sqrt{2\pi}} \; \int_{0}^{r^2/4 \lambda} \frac{{\rm e}^{-y/2}  \; {\rm d}y}{\sqrt{y \; (r^2- 4\Lambda y)}} \; 
\label{G2}
\end{eqnarray}

%Note the above result is derived 

\vspace{0.1in}

\noindent{\it Nearest-neighbor spacing distribution $P(s)$}: For a $2\times 2$ BE between Poisson and GOE, the probability of its nearest-neighbor eigenvalues  to occur at a distance $s$  can be given as \cite{ks}:
\begin{eqnarray}
P(s,\Lambda)=\frac{s}{4\Lambda} \; {\rm exp}\left(-\frac{s^2}{8  \Lambda}\right) 
\; \int_0^{\infty} {\rm d}x \; {\rm exp}\left[-\frac{x^2}{8\Lambda} -x \right] \; I_0 \left(\frac{x s}{4\Lambda} \right)  
\label{ps2}
\end{eqnarray}
with $I_0$ as the modified Bessel function. As $P(s)$ is dominated by the nearest neighbor pairs of eigenvalues, this result is expected to be 
a good  approximation for  $N \times N$ case too, especially in small-$s$ and small-$\Lambda$-result. This is confirmed by the perturbation theory based calculations for a general $N \times N$ matrix \cite{ks}.

\vspace{0.1in}

 \noindent{\it Number variance $\Sigma_2(r)$ and compressibility $\chi$}: 
The variance $\Sigma_2(r)$ of the number of the eigenvalues  in a range of $r$ mean level spacings,  is a measure of the long-range correlations in the spectrum. As discussed in \cite{me}, $\Sigma_2(r)$ can be written in terms of $R_2(r)$:  $\Sigma_2(r) = r -2\int_0^r (r-s)\; (1-R_2(s)) \;{\rm d}s$. The number variance for $r \gg \sqrt{\Lambda}$, within perturbation theory approach, can be given as \cite{fkpt} 
%(using eq.(\ref{g2}) ) 
\begin{eqnarray}
\Sigma_2(r,\Lambda) =r - 2\Lambda \left(\ln\frac{r^2}{2\Lambda} + \gamma -1+\ln 4 \right)
\end{eqnarray} 
with $\gamma$ as Euler's constant. 

For critical statistics analysis, it is more instructive to consider 
the level compressibility $\chi(r) =\frac{{\rm d} \Sigma_2(r)} {{\rm d}r} =1-2 \int_0^r (1-R_2(r)) \; {\rm d}r$, a measure of long-range rigidity of the spectrum .  A fractional value of $\chi(r)$ in $\lim_{r \rightarrow \infty, N \rightarrow \infty}$ is believed to be an indicator of the critical spectral statistics ($\chi \rightarrow 0$ for a GOE, $\chi=1$ for Poisson).
It is also related to the tail of $P(s)$: $\lim_{s \to \infty} \;  P(s) \; \sim \; {\rm e}^{-\frac{s}{ 2 \chi}}$. As discussed in \cite{psand}, for BEs intermediate to Poisson to Wigner transition, $\chi \approx (\pi^2 \Lambda)^{-1}$ for large $\Lambda$-cases. 

\section{Derivation of eq.(\ref{rnbc1})}

Consider the JPDF of the eigenvalues $P_{r}(r_1,\ldots, r_N)$ of a $N \times N$  ensemble with eigenvalues $r_1, \ldots, r_N$. The JPDF of the sub-ensemble consisting of  matrices with one non-random eigenvalue can be written as
\begin{eqnarray}
{\tilde P}_{r}(r_1,\ldots, r_N) = N \; P_{r}(r_1,\ldots, r_N) \; \delta(r_N)
\label{c0}
\end{eqnarray} 

The ensemble average of the level density $\rho(r)=\sum_{n=1}^N \delta(r-r_n)$ for the sub-ensemble can now be written as (with ${\rm D}^N r \equiv \prod_{k=1}^N {\rm d} r_k$)
 
\begin{eqnarray}
{\tilde R}_1(r) \equiv \langle \rho(r) \rangle &=&  \sum_{n=1}^N  \int  \; \delta(r-r_n) \; {\tilde P}_{r}(r_1,\ldots, r_N) \; {\rm D}^N r \nonumber \\
&=& N \; \sum_{n=1}^N  \int \delta(r-r_n) \; P_{r}(r_1,\ldots, r_{N}) \;  \delta(r_N)  \; {\rm D}^N r
\label{c1}
\end{eqnarray}
Separating the $N^{th}$ term from the rest, eq.(\ref{c1}) can be reduced as 
\begin{eqnarray}
{\tilde R}_1(r) 
&=& N \; \sum_{n=1}^{N-1}  \int \delta(r-r_n) \;  P_{r}(r_1,\ldots, r_{N-1}, 0) \; {\rm D}^{N-1} r + N \; \delta(r)  \int  \; P_{r}(r_1,\ldots,r_{N-1},r) \;   \; {\rm D}^{N-1} r \nonumber \\
&=& N (N-1)  \int  P_{r}(r_1,\ldots, r_{N-2}, r, 0) \; {\rm D}^{N-2} r + N \; \delta(r)  \int  \; P_{r}(r_1,\ldots,r_{N-1},r) \;   \; {\rm D}^{N-1} r \nonumber \\
\end{eqnarray}
Now using eq.(\ref{r02}) for $n=1, 2$ in the above equation, one has
\begin{eqnarray}
{\tilde R}_1(r)
&=&  R_2(r,0) +\delta(r) \; R_1(r)) \approx (1+\delta(r)) \; R_1(r)
\end{eqnarray}
Note here $R_2(r,0)$ and $R_1(r)$ correspond to the correlations of the full ensemble.  

Next we  derive eq.(\ref{rnbc1}) for $n=2$:

\begin{eqnarray}
{\tilde R}_2(r, r') \equiv \langle \rho_N(r) \rho_N(r') \rangle &=& \sum_{n,m=1}^N  \int  \; \delta(r-r_n) \; \delta(r'-r_m) \; {\tilde P}_{r}(r_1,\ldots, r_N) \; {\rm D}^N r
\end{eqnarray}
Again separating the $N^{th}$ contribution in the sum from the other terms, and using eq.(\ref{c0}), one can write
\begin{eqnarray}
{\tilde R}_2(r, r') 
&=& N \sum_{n,m=1}^{N-1}  \int \delta(r-r_n) \; \delta(r-r_n) \; P_{r} (r_1,\ldots, r_{N-1}, 0) \; {\rm D}^{N-1} r + \nonumber \\
&+& N\delta(r) \;\sum_{m=1}^{N-1}  \int \delta(r'-r_m) \;  P_{r}(r_1,\ldots, r_{N-1}, r) \; {\rm D}^{N-1} r + \nonumber \\
&+& N \delta(r') \;\sum_{n=1}^{N-1}  \int \delta(r-r_n) \;  P_{r}(r_1,..., r_{N-1}, r') \; {\rm D}^{N-1}r  \nonumber \\
&+& N\delta(r) \; \delta(r') \;\int  \; P_{r}(r_1,...,r_{N-1},r) \;   \; {\rm D}^{N-1} r
\label{c3}
\end{eqnarray}
Again using eq.(\ref{r02}) for $n=1, 2,3$ in the above equation, we get
%$R_2(r,r') =\sum_{n,m=1}^{N-1}  \int \delta(r-r_n) \; \delta(r-r_n) \; P_{r(N-1)}  \; {\rm D}^{N-1} r$.
\begin{eqnarray}
{\tilde R}_2(r, r') 
&=&  R_3(r,r',0) + \delta(r) \; R_2(r,r') +\delta(r') \;R_2(r,r') + \delta(r) \; \delta(r') \; R_2(r,r')  \\
& \approx & \left(1 + \delta(r)  +\delta(r')  + \delta(r) \; \delta(r') \right) \; R_2(r,r')
\label{c5}
\end{eqnarray}
 
Following similar steps, eq.(\ref{rnbc2}) can be derived for other $n$-values.


\begin{thebibliography}{10}

\bibitem{kirk}  S. Kirkpatrick, Rev. Mod. Phys. 45,  574, (1973). 

\bibitem{gc} V. Gurarie and J.T. Chalker, Phys. Rev. B, 68, 134207, (2003).
 
\bibitem{bp} Y.M.Beltukov and D.A.Parashin, JETP Letters, 93, 598, (2011).

%arXiv: 1011.2955v2.

\bibitem{bcm} B. Bellazzini, P. Calabrese and M. Mintchev, 
Phys. Rev. B, 79, 085122 (2009).

{
\bibitem{mm} M. Mezard, G. Parisi and A. Zee, Nucl. Phys. B, 559, 689, (1999).

\bibitem{gmpv} T.S. Grigera , V. Martin-Mayor, G. Parisi and P. Verrocchio, Phys. rev. Lett. 87, 085502, (2001); J. Phys. Cond. Mat. 14, 2167 (2002); Nature 422, 289 (2003). 

\bibitem{ps} F. A. Pinheiro and L. C. Sampaio, Phys. Rev. A 73, 013826, (2006).
\bibitem{scs} A. Svidzinsky, J-T Chang and M. O. Scully, Phys. Rev. A, 81, 053821, (2010).
\bibitem{rmo} M. Rusek, J. Mostowski and A. Orlowski, Phys. Rev. A, 61, 022704 (2000)
\bibitem{pin} F. A. Pinheiro, M. Rusek, A. Orlowski, and B. A. van Tiggelen, Phys. Rev. E 69, 026605, (2004).
\bibitem{ach} M. Antezza, Y. Castin and D. A. W Hutchinson, Phys. Rev. A, 82, 043602, (2010).
\bibitem{sm} S. Skipetrov and R. Maynard, Phys. Rev. Lett. 85, 736 (2000).
\bibitem{zs}  B. Zyuzin and A. Spivak, Phys. Rev. Lett. 84, 1970 (2000). 
\bibitem{gw} B. Gremaud and T. Wellens, Phys. Rev. Lett., 104, 133901 (2010).
\bibitem{gmpuv} T.S. Grigera , V. Martin-Mayor, G. Parisi , P. Urbani and P. Verrocchio , J.Stat. Mech., 110202015, (2011) ( arXiv :1011.2798)




\bibitem{os} C. R. Offer and B. D. Simons, J. Phys. A, 33, 7567, (2000). 

\bibitem{yvf} Y. V. Fyodorov, J. Phys. A: Math. Gen. 32, 7429, (1999).


\bibitem{luck}
J. Staring, B. Mehlig, Y.V.Fyodorov and J. M. Luck, Phys. Rev. E, 67, 047101, (2003). 

\bibitem{bray0} 
A. J. Bray and G. J. Rodgers, Phys. Rev. B, 38, 11461, (1988). 

}

\bibitem{adrs} A. Agarwal, S. das, S. rao and D. Sen, arXiv: 0810.3513v4.

\bibitem{shep}
B. Georgeot, O. Giraud, D.L. Shepelyansky, Phys. Rev. E 81, 056109, (2010).

\bibitem{cm} D. Challet and M. Marsilli, R. Zecchina, Phys. Rev. E, 84, 1824, (2000).




\bibitem{mar} S.R. Asmussen, "Markov Chains". Applied Probability and Queues. Stochastic Modelling and Applied Probability 51, (2003); G. Latouche and V. Ramaswami,  {\it Introduction to Matrix Analytic Methods in Stochastic Modeling}, 1st ed. , ASA SIAM, 1999.

\bibitem{me}
M.L.Mehta,{\it  Random Matrices}, Academic Press, (1991). 

\bibitem{stinch1}
R.B.Stinchcombe and I.R. Pimentel, Phys. Rev.B, 38, 4980, (1988). 


\bibitem{pbera}
Y. Sung Park and A.K. Bera,   "Maximum entropy autoregressive conditional heteroskedasticity model", Journal of Econometrics (Elsevier), 219, (2009).

\bibitem{itz}
A. Prats Ferrer, B. Eynard, P. Di. Francesco, J.-B. Zuber, J. Stat. Phys., 129, 885, (2007). 

\bibitem{gor}
T. Gorin, J. Math. Phys. 43, 3342, (2002). 



\bibitem{pich}
J-L. Pichard and B. Shapiro, J. Phys. I: France 4, 623, (1994).




\bibitem{dy}
F.Dyson, J. Math. Phys. 3, 1191 (1962).


\bibitem{ap}
A.Pandey, Chaos, Solitons and Fractals, 5, (1995).

\bibitem{fkpt}
J.B.French, V.K.B.Kota, A.Pandey and S.Tomsovic, Annals of Physics, 181, 198,(1988).

\bibitem{forres}
T. Nagao and P. J. Forrester, Physics Letters A 247, 42 (1998).


\bibitem{rp} 
N. Rosenzweig and C.E.Porter, Phys. Rev. 120, 1698 (1960). 

\bibitem{shapiro}
M. Krenin and B. Shapiro, Phys. Rev. Lett., 74, 4122, (1995); 
B. Shapiro, Int. J. Mod. Phys. B, 10, 3539, (1996).



\bibitem{alt} 
A. Altland, M. Janssen and B. Shapiro, Phys. Rev. E, 56, 1471, (1997).

\bibitem{ls}
F. Leyvraz and T.H. Seligman, J. Phys. A: Math. Gen. 23, 1555, (1990). 

\bibitem{ksh}
H.Kunz and B.Shapiro, Phys. Rev. E, 58, 400, (1998). 

\bibitem{fgm}
K.M.Frahm, T.Guhr, A.Muller-Groeling,  Ann. Phys. (N.Y.) 270, 292 (1998).  

\bibitem{gp}
T. Guhr and E. Papenbrock, Phys. Rev. E 59, 330, (1999).

\bibitem{hb}
 S. Hernandez-Quiroz and L. Benet, Phys. Rev. E 81, 036218, (2010). 

\bibitem{ks}
V.K.B.Kota and S.Sumedha, Phys. Rev. E, 60, 3405, (1999).
G.Lenz and F.Haake, Phys. Rev. Lett. 67, 1, (1991);
E.Caurier, B.Grammaticos and A. ramani, J.Phys. A, 23, 4903 (1990).
F.Leyvraz and T.H.Seligman, J. Phys. A, 23, 1555, (1990).
S.Tomsovic, Ph.D Thesis, University of Rochester (1986).

\bibitem{psc}
A. Pandey and P. Shukla, J. Phys. A, 24, 3907, (1991).

\bibitem{psand}
P.Shukla, Phys. Rev. E, 62, 2098, (2000);
J.Phys.: Condens. Matter 17, 1653, (2005).




\bibitem{ps-all}
P.Shukla, Phys. Rev. E, (71), 026226, (2005); Phys. Rev. E, 75, 051113, (2007); 
Phys. Rev. Lett., 87, 19, 194102, (2001).


\bibitem{psijmp}
P. Shukla, Int. J. Mod. Phys. B (WSPC) 26, 12300008, (2012). 

\bibitem{zirn}
A. Altland and M.R.Zirnbauer, Phys. Rev. B 55, 1142, (1997). 


\bibitem{ss}
S.Sadhukhan and P. Shukla, Random matrix ensembles with column/row constraints, part II. 

\bibitem{psr}
The universality and ergodicity are two separate concepts which may or may not  exist simultaneously. The universality of energy level-fluctuations for a class of systems implies that the fluctuations over the ensemble, representing any of these systems, at a given energy do not depend on any ensemble parameters and therefore system-dependent information (except for the global constraints representing the class) although they may vary from one energy to another. The ergodicity of spectral fluctuatons in random matrix ensembles implies that the fluctuations over an ensemble are same as the fluctuations along the spectral axis for a single matrix. An important example of universality without ergodicity is the Anderson Hamiltonian: the energy level fluctuations in an ensemble of Anderson Hamiltonians at a critical disorder are energy-dependent and non-ergodic (which is related to multifractality of wavefunctions at the critical point) but the level-statistics near $e=0$ is known to be universal. Another well-known example of universality without ergodicity is the level-statistics of integrable systems which is given by Poisson ensemble; a localization of wavefunctions makes these systems are non-ergodic.

%\bibitem{bray} 
%A. J. Bray and M.A.Moore, J. Phys. C: Sold State Phys., 14, 2629 (1981).


%\bibitem{shep2}
%B.V.Chirikov and D.L.Shepelyansky, Phys. Rev. Lett. 74, 518, (1995). 

%\bibitem{stinch}
%S.L.A. de Querioz and R.A. Stinchcombe, arXiv:cond-mat/0603043v1.



%\bibitem{ravin}
%R. Bhatt and S. Johri,  Int. J. Mod. Phys. Conf. Ser. 11, 79 (2012).

\bibitem{shep3}
D. L. Shepelyansky, Phys. Rev. Lett. 73, 2607, (1994);
Y.V. Fyodorov and A.D. Mirlin, Phys. Rev. B, 52, R11580, (1995).




% The critical point statistics for both of them  is same although the critical point appears in non-ergodic regime (partial localization of eigenfunctions). 

%In fact, a phase transition in general occurs due to some symmetry breaking which in turn corresponds to ergodicity breaking but the universality of statics at critical points is well-known. 





\end{thebibliography}
\end{document}